\documentclass{JHEP3} 
\pdfoutput=1
\usepackage{amssymb,bm,color}
\let\normalcolor\relax   % appears to be required when using color package! 
\let\ifpdf\relax % Looks like ifpdf is used by JHEP3, epsfig, and graphicx on my machine, and the three versions don't play nicely together. The next line is needed too, strangely.
\usepackage{ifpdf}
\usepackage{epsfig}
\usepackage{graphicx}
\usepackage{subfigure}
\usepackage{url}
\usepackage{amsmath}
\usepackage{amsbsy}

\def\Msun{M_\odot}
\def\hmpcinv{h\,{\rm Mpc}^{-1}}
\def\hinvgpc{h^{-1}{\rm Gpc}}
\def\hmpc{\,h^{-1}\,{\rm Mpc}}

\newcommand{\lmax}{\ell_{\rm max}}
\newcommand{\fsky}{f_{\rm sky}}

\newcommand{\fnl}{f_{\rm NL}}
\newcommand{\fnlk}{f_{\rm NL}(k)}

\newcommand{\perm}{{\rm perm.}}

\newcommand\ba{\begin{eqnarray}}
\newcommand\ea{\end{eqnarray}}
\newcommand{\FLSS}{F^{\rm LSS}}
\newcommand{\FCMB}{F^{\rm CMB}}
\newcommand{\kpiv}{k_{\rm piv}}
\newcommand{\nfnl}{{n_{\fnl}}}

\newcommand{\fnlstar}{f^*_{\rm NL}}

\newcommand{\zmax}{z_{\rm max}}

\newcommand{\fom}{{\rm FoM}^{\rm (NG)}}

\newcommand{\ells}[0]{
\ell_1 \ell_2 \ell_3
}

\newcommand{\eqn}[2]{
\begin{equation}\label{#1} 
#2 
\end{equation}
}

\title{Constraining Scale-Dependent Non-Gaussianity with Future
Large-Scale Structure and the CMB}

\author{Adam Becker${}^{1}$, Dragan Huterer${}^{1}$, Kenji Kadota${}^{1,2}$\\
{\small 
${}^{1}$ Department of Physics and Michigan Center for Theoretical Physics,
 University of Michigan, 450 Church Street, Ann Arbor, MI 48109; \\
${}^{2}$ Physics Department, Nagoya University, Nagoya, Aichi 464-8602, Japan}}

\abstract{ We forecast combined future constraints from the cosmic microwave
  background and large-scale structure on the models of primordial
  non-Gaussianity. We study the {\it generalized} local model of
  non-Gaussianity, where the parameter $\fnl$ is promoted to a function of
  scale, and present the principal component analysis applicable to an
  arbitrary form of $\fnlk$. We emphasize the complementarity between the CMB
  and LSS by using Planck, DES and BigBOSS surveys as examples, forecast
  constraints on the power-law $\fnlk$ model, and introduce the
    figure of merit for measurements of scale-dependent non-Gaussianity.}


\begin{document}

\maketitle

%%%%%%%%%%%%%%%%%%%%%%%%%%%%%%%%%%%%%%%%%%%%%%%
\section{Introduction}\label{sec:intro}

There has recently been a surge in interest to study departures in the
distribution of primordial density fluctuations from the random Gaussian case
predicted by standard inflationary models. The reason for this renewed
interest lies in the fact that any observable departures from Gaussianity
would essentially rule out the standard single-field, slow-roll inflationary
picture, pointing to a more complicated dynamics during the epoch of inflation
(see e.g.\ \cite{Chen_AA,Komatsu_CQG} for reviews).

It is therefore important to consider how one could parametrize primordial
non-Gaussianity. A much-studied model of primordial non-Gaussianity is the
local (or squeezed) model, which characterizes non-Gaussianity through a
single parameter $\fnl$ \cite{Salopek,Verde_CMBLSS,Komatsu_Spergel}
\eqn{eq:localNG}{
\Phi(x)=\phi_G(x)+\fnl(\phi_G(x)^2-\langle \phi_G (x)^2 \rangle ).
}
Here, $\Phi$ denotes the primordial curvature perturbations (Bardeen's
gauge-invariant potential), $\phi_G(x)$ is a Gaussian random field, and the
constant $\fnl$ is the parameter describing deviations from Gaussianity. The
local model has been much studied for its simplicity -- it contains the first
two terms of the most general local form of non-Gaussianity
\cite{Babich_shape}.
In a recent paper (\cite{Becker2011}; hereafter BHK11), we introduced a generalization of this
model to one where, in Fourier space, $\fnl=\fnl(k)$ is a function of scale
\begin{equation}
\Phi(k)=\phi_G(k)+\fnl(k)\int \frac{d^3 k'}{(2 \pi)^3}\phi_G(k')\phi_G(k-k').
\label{eq:fnlk_kspace}
\end{equation}
This is a natural extension of the popular 'local' model\footnote{Other
    models also display scale dependence of primordial non-Gaussianity; for example, the
    Dirac-Born-Infeld braneworld theory typically leads to scale-dependent
    equilateral $\fnl$, which has been constrained in Ref.~\cite{Bean_DBI}.}
$\fnl={\rm const}$, and it is physically well-motivated; it can describe
inflationary scenarios with multiple light fields, one of which is responsible
for the generation of curvature perturbations \cite{chris5,Shandera2010,Barnaby_dilaton}. The
bispectrum in this generalized local model is
\begin{equation}
B_{\phi}(k_1, k_2, k_3) = 2[\fnl(k_1) P_\phi(k_2)P_\phi(k_3) + \perm],
\label{eq:fnlk_bispec}
\end{equation}
%\
where $P_\phi$ is the power spectrum of potential fluctuations.  This reduces
to the familiar expression $B(k_1, k_2, k_3) = 2 \fnl (P_\phi(k_1)P_\phi(k_2)
+ \perm)$ when $\fnl$ is a constant. 

To parametrize this model while retaining its full generality, it is
convenient to consider a parametrization in piecewise-constant bins in
wavenumber:
\begin{equation}
\fnl^i\equiv \fnl(k_i)
\label{eq:fnl_piecewise}
\end{equation}
where each $\fnl^i$ is the value of $\fnl(k)$ in the $i$-th wavenumber bin. In
BHK11, we used this parametrization to project errors on $\fnl(k)$ from a
hypothetical Stage III galaxy survey. As in BHK11, we adopt 20 bins in
wavenumber distributed uniformly in $\log(k)$, which is easily sufficient to obtain
the best-measured principal components accurately.

In this work, we perform an analysis that is similar in spirit to that in
BHK11, but extended in several respects. First of all, we develop formalism
and work out forecasts for how well the CMB, in particular Planck
\cite{Planck_paper}, can measure $\fnlk$. We combine this with the LSS
forecasts updated to reflect three specific galaxy surveys. Having done that,
we obtain a clearer picture of where we can expect good constraints on
non-Gaussianity in $k$-space.  Finally, we project our forecasts to the
specific, power-law model in wavenumber, and thus clarify at which wavenumber
the CMB, LSS and the combined surveys best determine
non-Gaussianity. Therefore, this work complements not only BHK11 and studies
that forecasted errors for (occasionally slightly different) models of
scale-dependent non-Gaussianity
\cite{Sefusatti2009,Shandera2010,Giannantonio:2011ya}, but also many previous
forecasts for future constraints on {\it constant} $\fnl$ 
\cite{Sefusatti,Smith_Zaldarriaga,Yadav2007,McDonald,Carbone,Slosar:2008ta,Carbone2,Fergusson2010,Sartoris,Cunha_NG,Fedeli:2010ud,Joudaki,Giannantonio:2011ya,Pillepich:2011zz,Hazra:2012qz}.

The structure of this paper is as follows: in Sec.~\ref{sec:LSS}, we briefly
review the main result from BHK11 -- the signature of the generalized local
model on large-scale structure through halo bias -- and explore the effects of
an additional term in the modeling of non-Gaussian bias first pointed out by
Desjacques et al.\ \cite{Desjacques2011}. In Sec.~\ref{sec:CMB}, we find the
signature of the generalized local model on the CMB bispectrum, with
particular emphasis on Planck. Finally, in Sec.~\ref{sec:Results} we combine
the results for a set of joint constraints; we also perform a
principal-component analysis, and project constraints on a power-law model of
$\fnl(k)$. We conclude in Sec.~\ref{sec:concl}. Details of the computational
work can be found in the Appendices.

%%%%%%%%%%%%%%%%%%%%%%%%%%%%%%%%%%%%%%%%%%%%%%%
\section{Generalized local model: signatures in large-scale structure}
\label{sec:LSS}

In this section we first briefly review how the general local model of
non-Gaussianity affects the bias of dark matter halos. We then present details
of the LSS surveys that we will consider. 

\subsection{Effect on the bias}

For the local non-Gaussian model from Eq.~(\ref{eq:localNG}), the dark matter
halo bias acquires scale dependence \cite{Dalal}:
\begin{equation}
b(k)
= b_0 + \Delta b(k)
=b_0 + \fnl(b_0-1)\delta_c\, \frac{3\Omega_mH_0^2}{a\,g(a) T(k)c^2 k^2},
\label{eq:bias}
\end{equation}
where $b_0$ is the usual Gaussian bias (on large scales, where it is
constant), $\delta_c\approx 1.686$ is the collapse threshold, $a$ is the scale
factor, $\Omega_m$ is the matter density relative to the critical density,
$H_0$ is the Hubble constant, $k$ is the wavenumber, $T(k)$ is the transfer
function, and $g(a)$ is the growth suppression factor\footnote{The usual
  linear growth $D(a)$, normalized to be equal to $a$ in the matter-dominated
  epoch, is related to the suppression factor $g(a)$ via $D(a)=ag(a)/g(1)$,
  where $g(a)$ is normalized to be equal to unity deep in the matter-dominated
  epoch.}.  See also
\cite{MV,Grossi,Slosar_etal,Afshordi_Tolley,McDonald,Taruya08,Desjacques_Seljak_Iliev,PPH,Valageas:2009vn,GP,Cunha_NG,Schmidt_Kam,Desjacques_long,Wagner:2010me,Wagner:2011wx,Sefusatti:2011gt}
who explored the effects of primordial non-Gaussianity on the bias of dark
matter halos in great detail.

In BHK11, we worked out the signature of the generalized local model, using
the MLB formalism \cite{Grinstein:1986en,MLB1986,MV}, and made forecasts for
future galaxy surveys. It is not our intention to fully repeat all of our
analysis from that paper, and we just quote essential results. The change in
the bias $\Delta b$ is given, in the generalized local model, by Eq.~(3.16)
from BHK11
\begin{eqnarray}
{\Delta b\over b} (k)& =&
{\delta_c\over D(z)}\, {2\over 8\pi^2 \sigma_R^2 \mathcal{M}_R(k)}
\int dk_1 k_1^2 \mathcal{M}_R(k_1)P_\phi(k_1)
\nonumber \\[0.2cm]
&\times &\int d\mu \mathcal{M}_R(k_2) \left [ \fnl(k){P_\phi(k_2)\over
    P_\phi(k)} + 2\fnl(k_2)\right ].
\label{eq:deltab_b_old}
\end{eqnarray}
where $\mathcal{M}_R(k) \equiv [k^2 T(k)\tilde{W}_R(k)]/(H_0^2 \Omega_{m}) $
and $\tilde{W}_R(k)$ is the Fourier transform of the top-hat filter with
radius $R$. The derivatives with respect to piecewise constant parameters
$\fnl^i$ are straightforward and given by Eq.~(3.19) of our previous paper.

However Eq.~(\ref{eq:deltab_b_old}) is not entirely correct in describing the
scale-dependent bias from a general NG model. Desjacques et
al.\ (\cite{Desjacques2011}; see also \cite{Desjacques_long}) pointed out that
the expression \eqref{eq:bias} is only correct in the high-peak, small-$k$
limit.  An additional term is required for the exact expression:
\begin{eqnarray}
\Delta b(k)= \frac{2F(k)}{M_R(k)D(z)}\left [ (b_0 - 1)
  \delta_c+\frac{d \ln F(k)}{d \ln \sigma_{R}} \right ]
\label{eq:Desjacques}
\end{eqnarray}
where
\begin{eqnarray}
F(k) =
 {1\over 8\pi^2 \sigma_R^2}
\int dk_1 k_1^2 \mathcal{M}_R(k_1)P_\phi(k_1)
 \int^{1}_{-1} d\mu \mathcal{M}_R(k_2) \left [ \fnl(k){P_\phi(k_2)\over P_\phi(k)} + 2\fnl(k_2)\right ].
\end{eqnarray}

The new term (second term in square parentheses in Eq.~(\ref{eq:Desjacques}))
vanishes when the fiducial model for non-Gaussianity is $\fnl(k) = {\rm
  const}$, but it becomes relevant for truly scale-dependent models, including
the piecewise-constant parametrization of $\fnl(k)$ from equation
\eqref{eq:fnl_piecewise}. Because we are expanding (taking Fisher derivatives)
around the constant value of $\fnlk$ (30 or zero), we find very small
though nonzero effect of this new term describing halo bias. See Appendix
\ref{app:Desjaqcues} for details.

%%%%%%%%%%%%%%%%%%%%%%%%%%%%%%%%%%%%%%%%%%%%%%%
\subsection{Fisher matrix analysis: assumptions and survey specifications}
\label{sec:LSS_survey}

We are interested in making forecasts for constraints on non-Gaussianity from
future galaxy surveys, and for this we employ the standard Fisher matrix
formalism. 

For measurements of the power spectrum of dark matter halos, the Fisher matrix
$F$ is \cite{Tegmark97}
\begin{equation}
\FLSS_{ij} = \sum_m V_m \int_{k_{\rm min}}^{k_{\rm max}} 
\frac{\partial P_h(k, z_m)}{\partial p_i} \frac{\partial P_h(k, z_m)}{\partial
  p_j} 
\,\frac{1}{\left[ P_h(k, z_m) + \displaystyle\frac{1}{n} \right]^2}\, \frac{k^2 dk}{(2\pi)^2},
\label{eq:LSSFisher}
\end{equation}
where $V_m$ is the comoving volume of the $m$-th redshift bin, each redshift
bin is centered on $z_m$, and we have summed over all redshift bins.  We
adopt\footnote{We found that the constraints are insensitive to the
    precise value of $k_{\rm min}$ for the fiducial $\fnl(k) = 30$ (or any
    sufficiently nonzero value), since the most constraining scales are
    intermediate between $k_{\rm min}$ and $k_{\rm max}$, reflecting the
    competition between larger noise and larger signal as one goes to lower
    $k$. However, for the fiducial value $\fnl(k) = 0$ the constraints indeed come
    from the largest available scales, and in that case we adopt $k_{\rm min}
    = V_{\rm survey}^{-1/3}$.}   $k_{\rm min}=10^{-4}\hmpc$, and we choose
$k_{\rm max}$ as a function of $z$ so that $\sigma(\pi/(2k_{\rm max}), z) =
0.5$ \cite{SeoEisenstein2003}, which leads to $k_{\rm max}(z= 0) \approx
0.1\hmpcinv$. $P_h$ is the dark matter halo power spectrum, related to the
true dark matter power spectrum $P$ through
\begin{equation}
P_h(k)=b(k)^2 P(k),
\label{eq:bhalo}
\end{equation}
where each quantity implicitly also depends on redshift. Finally, $p_{i}$ are
the parameters of interest; in our case, these are the $\fnl^i$, cosmological
parameters, and the bias-related nuisance parameters listed below in
\eqref{eq:params}. The minimal error in the $i$-th cosmological parameter is,
by the Cram\'er-Rao inequality, $\sigma(p_i)\simeq \sqrt{(F^{-1})_{ii}}$.

All of the results cited in this section, as well as LSS survey projections
elsewhere in this paper, assume the following survey properties, modeled on
BigBOSS \cite{BigBOSS_paper}, unless explicitly stated otherwise. We adopt the
fiducial value $\fnl = 30$, chosen to roughly correspond to the
maximum-likelihood value favored by current CMB data \cite{wmap7}. The
fiducial cosmological model is the standard $\Lambda$CDM model with Hubble's
constant $H_0$; physical dark matter and baryon densities $\Omega_{\rm cdm}
h^2$ and $\Omega_{\rm b} h^2$; equation of state of dark energy $w$; the log
of the scalar amplitude of the matter power spectrum, $\log A_s$; and the
spectral index of the matter power spectrum, $n_s$ . Fiducial values of these
parameters correspond to their best-fit WMAP7 values \cite{wmap7}.  We also
added the forecasted cosmological parameter constraints from the CMB
experiment Planck by adding its Fisher matrix as a prior (W.\ Hu, private
communication). Note that the CMB prior does {\it not} include CMB constraints
on non-Gaussianity, so that we are not double-counting the latter constraints.
Next, we include twenty piecewise-constant non-Gaussianity parameters
$\fnl(k_i)\equiv \fnl^i$ with $i=0, 1, \ldots, 19$. Finally, we include a
Gaussian bias parameter in our Fisher matrix, $b_0(z)$, for each of the 44
redshift bins over the range $0.1 < z < 4.5$ (which is discussed further
below). The full list of parameters that we have in the BigBOSS Fisher matrix
is
\begin{equation}
\{ p_i\} = \{H_0, \Omega_{\rm cdm} h^2, \Omega_{\rm b} h^2, w, \log A_s, n_s, b_0^1, \ldots, b_0^{44}, \fnl^0, \ldots, \fnl^{19}\}
\label{eq:params}
\end{equation}
BigBOSS will utilize three different tracers: LRGs, ELGs, and QSOs. The
fiducial values for $b_0(z = 0)$ are different for each of the three tracers,
as are the number densities. To account for this, we calculated the Fisher
matrix for each tracer, then added all three together for the final BigBOSS
Fisher matrix. The fiducial values for $b_0(z = 0)$ were 1.7 for the LRGs,
0.84 for the ELGs, and 1.2 for the QSOs \cite{BigBOSS_internal}. In each case,
we assume the simple scaling of $b_0$ with redshift, $b_0(z) = b_0(z=0)/D(z)$.

In order to simplify the analysis, and in light of the uncertainties of the
distribution of {\it observed} LSS tracers' masses, we assume a fixed halo
mass of $10^{13} \Msun / h$.  Since we marginalize over $b_0$ in each redshift
bin, we effectively delete much information about the mass of the
tracers. Essentially, we utilize information about the redshift- and
wavenumber-dependence of bias, but avoid --- at least for now --- using
information about the masses, since accurate masses of LSS tracers are
typically very difficult to obtain, except for galaxy clusters.

We assume that BigBOSS will cover 14,000 square degrees. The redshift ranges
for the LRGs, ELGs and QSOs are $z^{\rm LRG}_{\rm max} = 1.2$, $z^{\rm
 ELG}_{\rm max} = 1.8$ and $z^{\rm QSO}_{\rm max} = 4.5$; the forecasted
number density in each redshift bin of course widely vary, and in particular
the QSOs have much lower typical number densities per redshift bin than the
other tracers \cite{BigBOSS_internal}. We split the survey into the
aforementioned 44 redshift bins out to $z_{\rm max}=4.5$; the total volume of
the survey is therefore $V_{\rm tot}=230\,(\hinvgpc)^3$.  The errors in the
cosmological parameters vary, in the cosmic variance limit, as $V_{\rm
  tot}^{-1/2}$.

In addition to presenting our results with the fiducial BigBOSS survey (14,000
sq.\ deg., $\zmax=4.5$), we also forecast constraints for the Dark Energy
Survey (DES \cite{DES_paper}),
which will cover 5,000 sq.\ deg.\ with $\zmax = 1.0$, split into 5 redshift bins. For
simplicity's sake, we assume that this survey will only see one tracer, with a
fiducial Gaussian bias $b_0(z = 0) = 2.0$ and a number density $n = 2 \times
10^{-4}\, (\hmpcinv)^3$ (independent of redshift). Note that the actual
redshift distribution and number density of the DES tracers will depend on the
{\it spectroscopic followup} to the main survey and the objects that it
targets, details of which are somewhat uncertain at this time and hence our
approximate assumptions.  The rest of the Fisher matrix formalism for this
survey (cosmological parameters, etc.) is the same as what we used for the
BigBOSS Fisher matrix.

%%%%%%%%%%%%%%%%%%%%%%%%%%%%%%%%%%%%%%%%%%%%%%%
\subsection{The effect of the fiducial value on constraints}
\label{sec:fidval}

\begin{figure}[t] %htbp, h = here, t = top, b = bottom, p = page: if nothing there...
\begin{center}
\includegraphics[width= 6in]{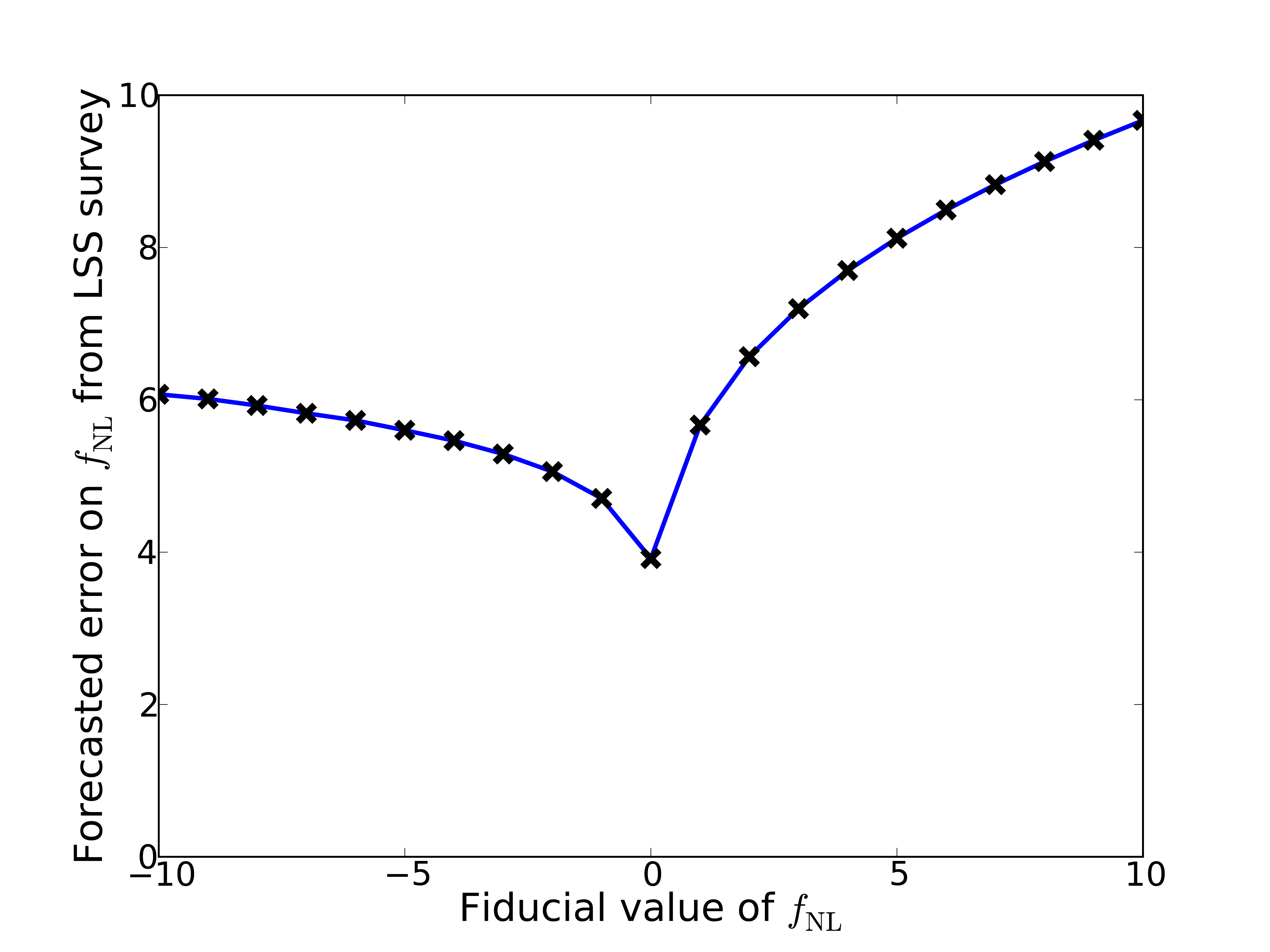}
\caption{ A more detailed look at how the choice of fiducial $\fnl$ affects
  the projected constraints on {\it constant} $\fnl$ from a future galaxy
  survey.  See text for analytic explanation for why results are the best at
  fiducial value of $\fnl=0$. }
% NB: You need to put a \protect on your references in the captions, otherwise LaTeX has a hissy fit and says that there's an "extra }" here.
\label{fig:lss_fiducial_local}
\end{center}
\end{figure}

Looking back at Equation \eqref{eq:LSSFisher}, we see that the fiducial $\fnl$
enters through the bias, by way of $P_h$. Assuming $P_h(k) \gg 1/n$ (a
reasonable assumption  at large angular scales where non-Gaussianity
constraints largely come from and where shot noise is negligible), we find
that the Fisher matrix element corresponding to $\fnl={\rm const}$ is
\begin{equation}
\FLSS \propto \int \left(\frac{\partial b(k)}{\partial \fnl}\right)^{2}{b^{-2}(k)} dk
= \int \left ( {\Delta b(k) \over \fnl \left( b_0 + \Delta b(k) \right ) } \right)^2 dk
%= \int \left ( \frac{3\Omega_mH_0^2}{a\,g(a) T(k)c^2 k^2} {(b_0-1)\delta_c \over \left( b_0 + \Delta b(k) \right ) } \right)^2 dk
\label{eq:F_fnl_const}
\end{equation}
Thus, the expression on the right-hand side will, in general, be dependent on the
choice of fiducial $\fnl$. 
Since $|\Delta b(k)|$ blows up at small $k$, in that regime we have:
\begin{equation}
\left ( {\Delta b(k) \over \fnl \left( b_0 + \Delta b(k) \right ) } \right)^2 \approx {1 \over \fnl^2}
\end{equation}
At large $k$, $\Delta b(k)$ goes to 0, taking the entire expression with
it. Thus, the integral is dominated by the contribution at low $k$, meaning we
should expect a maximal Fisher matrix element around a fiducial $\fnl = 0$.
And indeed, that is what we see in Figure \ref{fig:lss_fiducial_local}: the
projected constraints on $\fnl$ from a given sky survey depend strongly on the
fiducial value chosen, with the tightest constraints at $\fnl = 0$.

We note that, as shown in the following section, the Fisher matrix is
independent of the fiducial $\fnl$ value for the CMB constraints for our
piecewise-constant parameterization.

%%%%%%%%%%%%%%%%%%%%%%%%%%%%%%%%%%%%%%%%%%%%%%%
\section{Generalized local model: signatures in the CMB}
\label{sec:CMB}

Traditionally, the best constraints on non-Gaussianity have come from the
CMB. This is done almost exclusively through estimators involving the N-point
correlation functions for $N > 2$ and their Fourier transforms, the
polyspectra. Most emphasis has been on the $N = 3$ case, or the bispectrum of
temperature fluctuations in the CMB, if only because of its relative
computational simplicity.  The well-known general expression for the CMB
bispectrum, re-derived in Appendix \ref{app:bispec}, is
\begin{eqnarray}
B^{pqr}_{\ell_1 \ell_2 \ell_3} &=& 
\left({2 \over \pi}\right)^3
\sqrt{(2 \ell_1 + 1) (2 \ell_2 + 1) (2 \ell_3 + 1) \over 4 \pi}
\binom{\ell_1 \; \ell_2 \; \ell_3}{0 \;\; 0 \;\; 0}
\int k^2_1 d k_1\: k^2_2 d k_2 \: k^2_3 d k_3 \nonumber \\[0.2cm]
&\times & B_{\Phi} (k_1, k_2, k_3) t^p_{\ell_1}(k_1) t^q_{\ell_2}(k_2) t^r_{\ell_3}(k_3) \int_0^{\infty} r^2 d r \;  j_{\ell_1}(k_1 r)
j_{\ell_2}(k_2 r)
j_{\ell_3}(k_3 r)
\label{eq:GeneralCMBBispectrum}
\end{eqnarray}
where the expression in angular parentheses is the Wigner-3j symbol, $B_\Phi$
is the curvature bispectrum, and $t_\ell$ are the radiation transfer functions.

In principle, we can use this to find the Fisher matrix $F_{ij}$ for the CMB
bispectrum and thus forecast how well the CMB bispectrum can determine the
cosmological parameters:
 \cite{YadavWandelt2010, BabichZal2004, Komatsu_Spergel, SpergelGold1999}%
\begin{equation}
\FCMB_{ij} = \fsky \sum_{lmn, pqr} \sum_{2 \leq \ell_1 \leq \ell_2 \leq \ell_3} 
{1 \over \Delta_{\ells} } {\partial B^{lmn}_{\ells} \over \partial p_i} 
(\mathbf{C}_{\ells}^{-1})_{lmn, pqr} {\partial B^{pqr}_{\ells} \over \partial p_j}.
\label{eq:BiFisher}
\end{equation}
Here, $\mathbf{C}$ is the covariance of the bispectra and $p_{i,j}$ are the
parameters of interest. $\Delta_{\ells}$ is a combinatoric term -- equal to 6
when $\ell_1 = \ell_2 = \ell_3$, 1 when $\ell_1 \neq \ell_2 \neq \ell_3$, and
2 otherwise \cite{SpergelGold1999}. The indices $i,j,k$ and $p,q,r$ run
independently over all eight possible ordered triplets of temperature and
polarization fields (TTT, TTE\dots EEE). The details of calculating
$B^{pqr}_{\ells}$ and its derivatives are in Appendix \ref{app:bispec}, while
the details of calculating the bispectrum covariance $\mathbf{C}$ are in
Appendix \ref{app:covariance}.

%%%%%%%%%%%%%%%%%%%%%%%%%%%%%%%%%%%%%%%%%%%%%%%
%\subsubsection{The local model}
%\label{sec:localmodel}
Equation \eqref{eq:GeneralCMBBispectrum} is a totally general result for the
bispectrum of the CMB in terms of the Bardeen curvature bispectrum; we have
not picked any model of non-Gaussianity. But
\eqref{eq:GeneralCMBBispectrum} is not useful without picking a form for
$B_{\Phi} (k_1, k_2, k_3)$.
For the constant $f_{\rm NL}$ case, we have the following Bardeen curvature bispectrum:
\begin{equation}
B_{\Phi} (k_1, k_2, k_3) = 2 \Delta^2_{\phi} f_{\rm NL} \left( {1 \over k_1^{3 - (n_s - 1)}  k_2^{3 - (n_s - 1)} } + \mbox{perm.} \right)
\label{eq:ConstBispec}
\end{equation}
where $\Delta_{\phi}$ is the amplitude of the curvature power spectrum.  Using
Eqs.~\eqref{eq:BiFisher}, \eqref{covMF}, and \eqref{ConstCMBbispecMF}, we have
%This yields 
the following expression for the CMB bispectrum Fisher information in the constant $\fnl$ case:
\begin{eqnarray}
\FCMB_{\fnl} 
& = &
4 \Delta^4_{\phi} \sum_{lmn, pqr} 
\sum_{2 \leq \ell_1 \leq \ell_2 \leq \ell_3} {1 \over \Delta_{\ells}}
{(2 \ell_1 + 1) (2 \ell_2 + 1) (2 \ell_3 + 1) \over 4 \pi} \binom{\ell_1 \; \ell_2 \; \ell_3}{0 \;\; 0 \;\; 0}^2 
{1 \over \Delta_{\ells}}   \nonumber
\\ [0.2cm] 
& \times &(C^{-1}_{\ell_1})_{lp} (C^{-1}_{\ell_2})_{mq} (C^{-1}_{\ell_3})_{nr} 
 \left[ \int_0^{\infty} r^2 d r \left( \alpha^l_{\ell_1}(r)
   \beta^m_{\ell_2}(r)  \beta^n_{\ell_3}(r) + \mbox{perm.} \right) \right] 
\label{eq:ConstFisher}
\\[0.2cm]  
& \times &
\left[ \int_0^{\infty} r^2 d r \left( \alpha^p_{\ell_1}(r) \beta^q_{\ell_2}(r) 
\beta^r_{\ell_3}(r) + \mbox{perm.} \right) \right]\nonumber
\end{eqnarray}
where $\alpha_\ell$ and $\beta_\ell$ are defined in equations \eqref{alphaMF}
and \eqref{betaMF}.

For the {\it scale-dependent} $\fnl(k)$ case from our generalized ansatz,
things are somewhat more complicated. The Bardeen curvature bispectrum is:
\begin{equation}
B_{\Phi} (k_1, k_2, k_3) = 2 \Delta^2_{\phi} \left( {\fnl(k_3) \over k_1^{3 - (n_s - 1)}  k_2^{3 - (n_s - 1)} } + \mbox{perm.} \right).
\label{eq:fnlkBispec}
\end{equation}
Using the piecewise-constant parametrization of $\fnl(k)$,
Eqs.~\eqref{eq:BiFisher}, \eqref{covMF}, and \eqref{fnlkDerivMF} yield the
following expression for the Fisher matrix of all the $\fnl^i$ in the
scale-dependent case, similar to Eq.~(\ref{eq:ConstFisher}):

\begin{eqnarray}
\FCMB_{ij} 
& =  &
4 \Delta^4_{\phi} \sum_{lmn, pqr}
\sum^{\lmax}_{2 \leq \ell_1 \leq \ell_2 \leq \ell_3} {1 \over \Delta_{\ells}}
{(2 \ell_1 + 1) (2 \ell_2 + 1) (2 \ell_3 + 1) \over 4 \pi}
\binom{\ell_1 \; \ell_2 \; \ell_3}{0 \;\; 0 \;\; 0}^2 
\nonumber  \\[0.2cm]
& \times &
{1 \over \Delta_{\ells}} (C^{-1}_{\ell_1})_{ip} (C^{-1}_{\ell_2})_{jq} (C^{-1}_{\ell_3})_{kr} 
\left [ \int_0^{\infty} r^2 d r \left( \alpha^{l,i}_{\ell_1}(r) \beta^m_{\ell_2}(r) \beta^n_{\ell_3}(r) + \mbox{perm.} \right) \right]
 \\[0.2cm]
 & \times & 
\left [ \int_0^{\infty} r^2 d r \left( \alpha^{p,j}_{\ell_1}(r) \beta^q_{\ell_2}(r) \beta^r_{\ell_3}(r) + \mbox{perm.} \right) \right].
\nonumber 
\label{eq:fnlkFisher}
\end{eqnarray}
Despite appearances, this is a relatively straightforward calculation to
perform, and it takes roughly an hour (on four processors) for twenty $\fnl^i$
parameters with $\lmax \approx 2000$.

We did not include other cosmological parameters in the CMB bispectrum Fisher
matrix, since the bispectrum does not constrain them terribly well, while on
the other hand the CMB {\it power} spectrum places very good constraints on
the other cosmological parameters. In other words, non-Gaussianity estimates
obtained using the CMB bispectrum would not be significantly affected by
marginalizing over the other cosmological parameters within their allowed
ranges, as explicitly shown by Ref.~\cite{Liguori_Riotto}. Therefore, it is a
  fair (and certainly very helpful) approximation to think the CMB power
  spectrum and the bispectrum complementing each other by constraining the
  standard cosmological parameters, and the non-Gaussian parameters,
  respectively and separately. This has indeed been the approach in the
  literature (e.g.\ \cite{Komatsu_Spergel, BabichZal2004}).

%%%%%%%%%%%%%%%%%%%%%%%%%%%%%%%%%%%%%%%%%%%%%%%
\section{Results and Joint Constraints}
\label{sec:Results}

\begin{figure}[t] 
\begin{center}
\subfigure[BigBOSS]{\includegraphics[width= 2.9in]{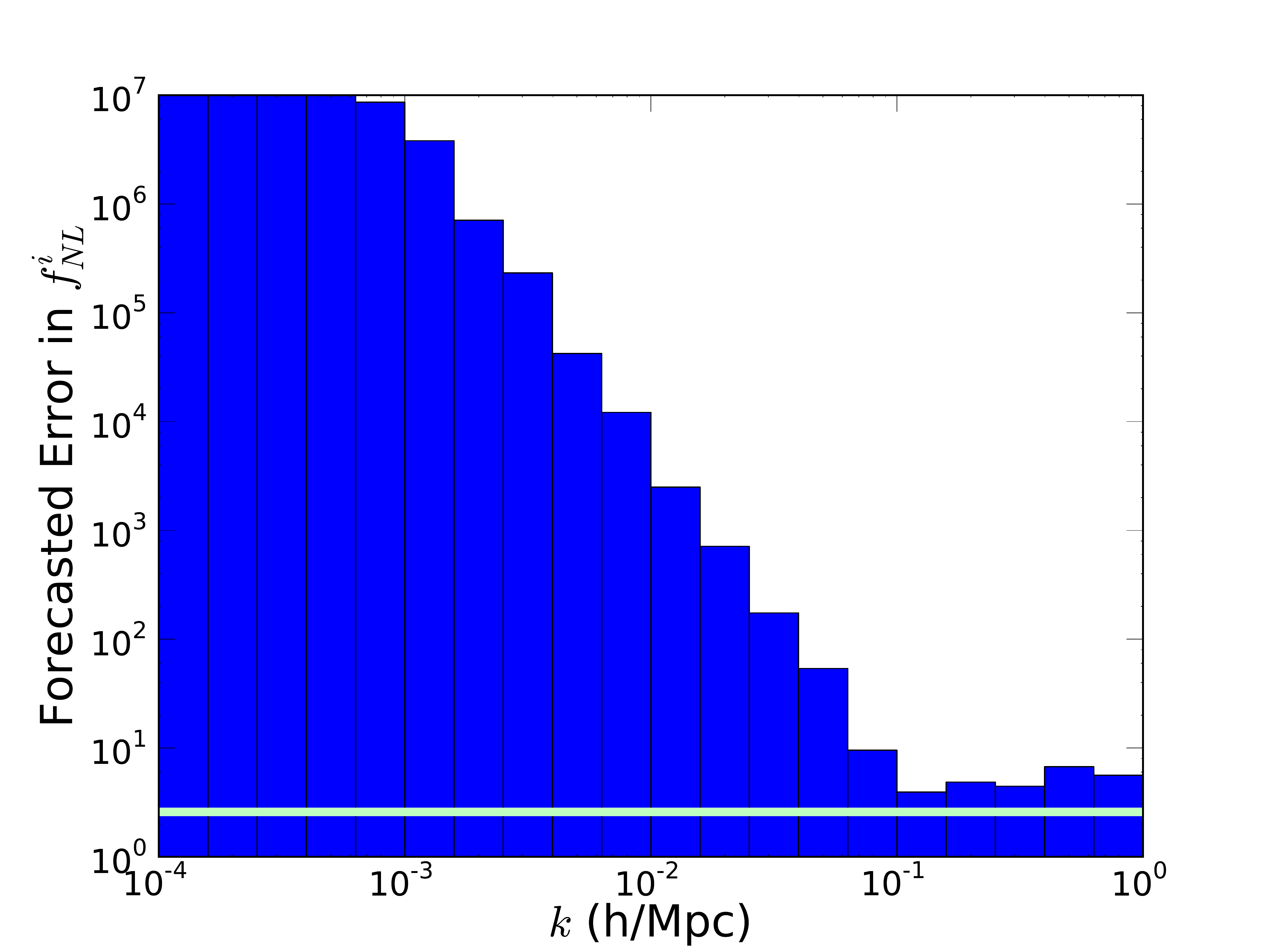}}
\subfigure[Planck]{\includegraphics[width= 2.9in]{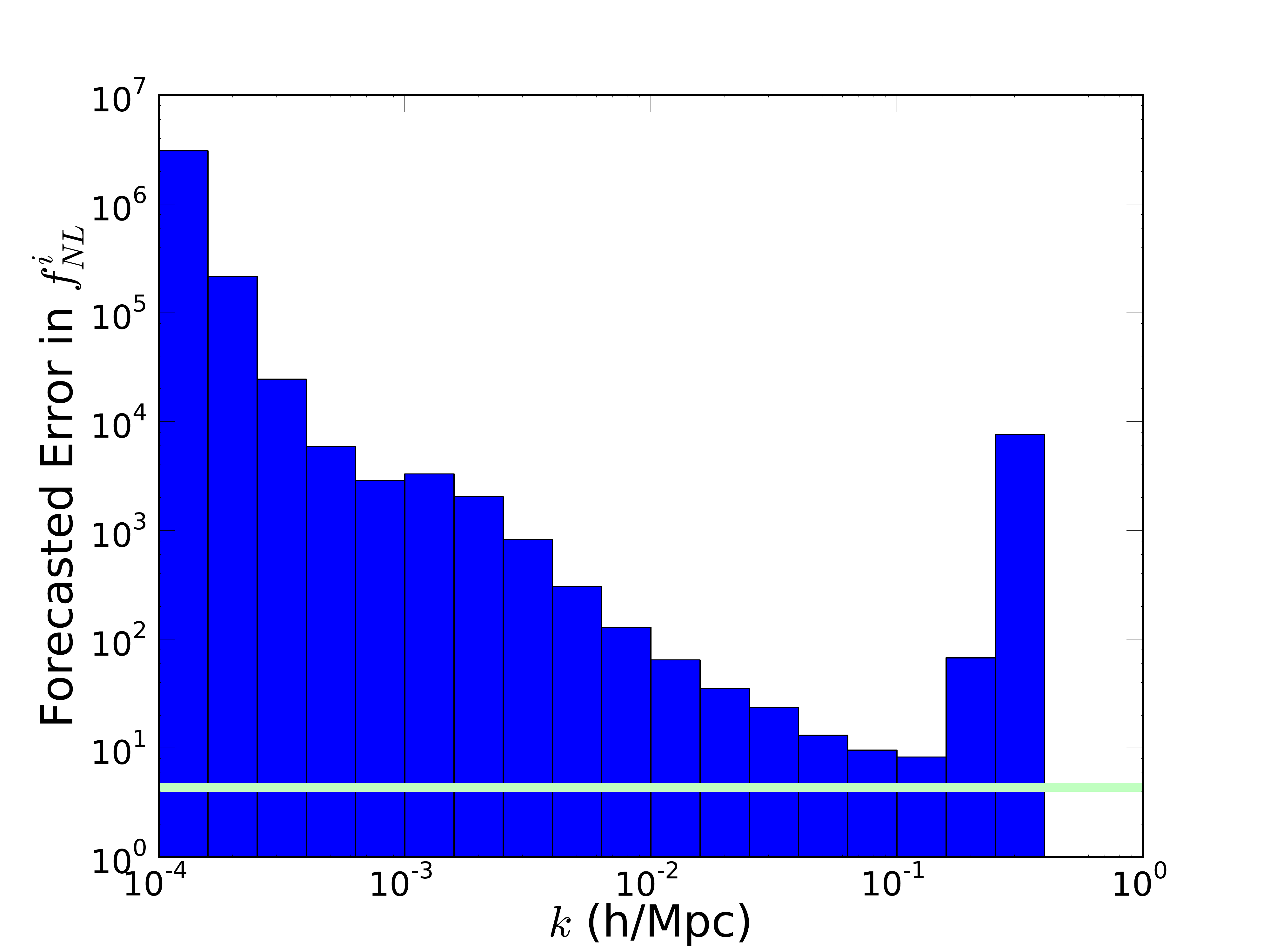}}
	\subfigure[Combined]{\includegraphics[width = 2.9in]{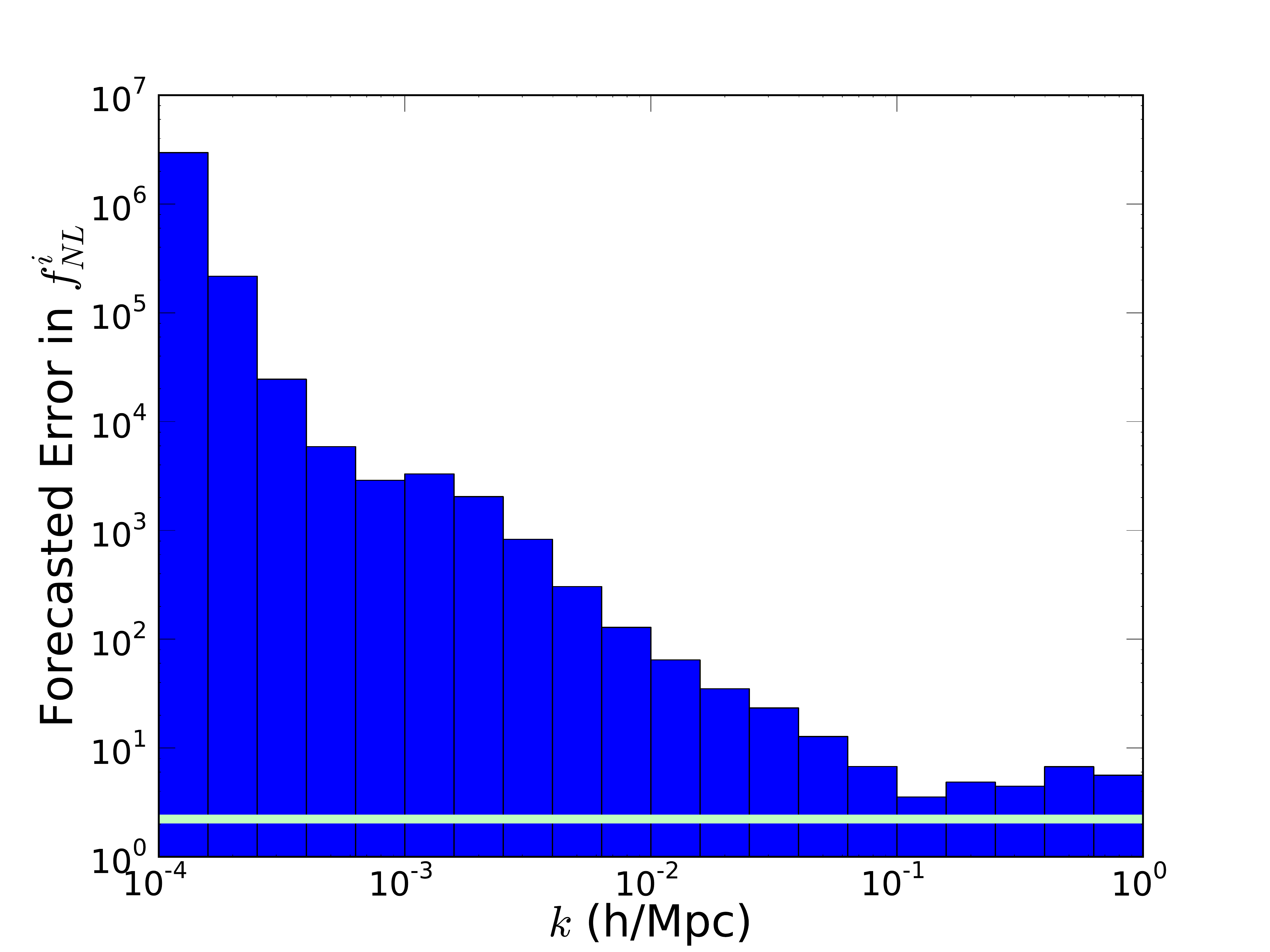}}
        \caption{Constraints on the piecewise constant parameters $\fnl^i$ in
          the generalized local model with the LSS (top left), CMB (top
          right), and LSS+CMB (bottom). All constraints are unmarginalized, in
          order to more clearly show the wavenumber-dependent sensitivity of
          the probes to primordial non-Gaussianity. The LSS constraints come
          from the power spectrum of galaxies, while the CMB constraints come
          from the bispectrum of temperature fluctuations. See text for
          details. For reference, the green line is the constraint found for a
          constant $\fnl$ using the same assumptions. There are bins
          ``missing" on the rightmost end of the Planck plot; those bins
          correspond to $k$-values too large to be probed when $\ell_{\rm max}
          = 2000$, as it is here.
% NB: You need to put a \protect on your references in the captions, otherwise LaTeX has a hissy fit and says that there's an "extra }" here.
}
\label{fig:unmarg_errs}
\end{center}
\end{figure}

\begin{figure}[t]
\begin{center}
\includegraphics[width=2.9in]{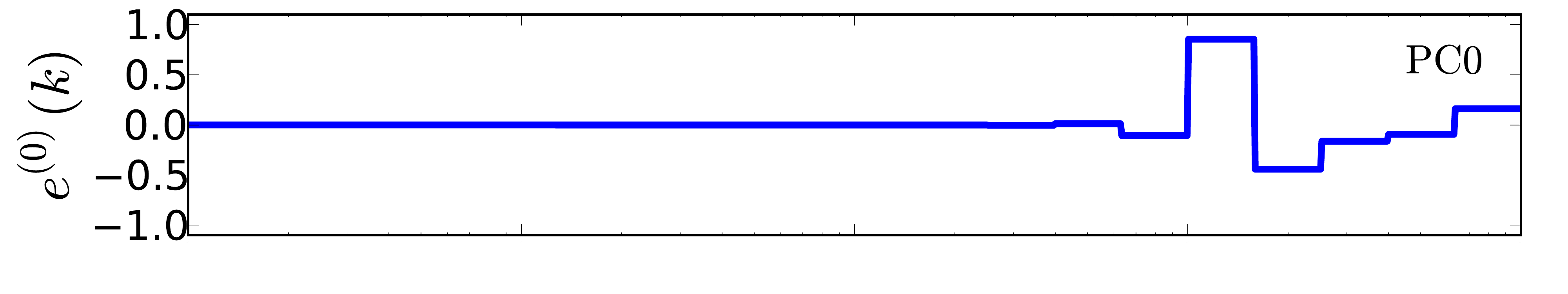} \includegraphics[width=2.9in]{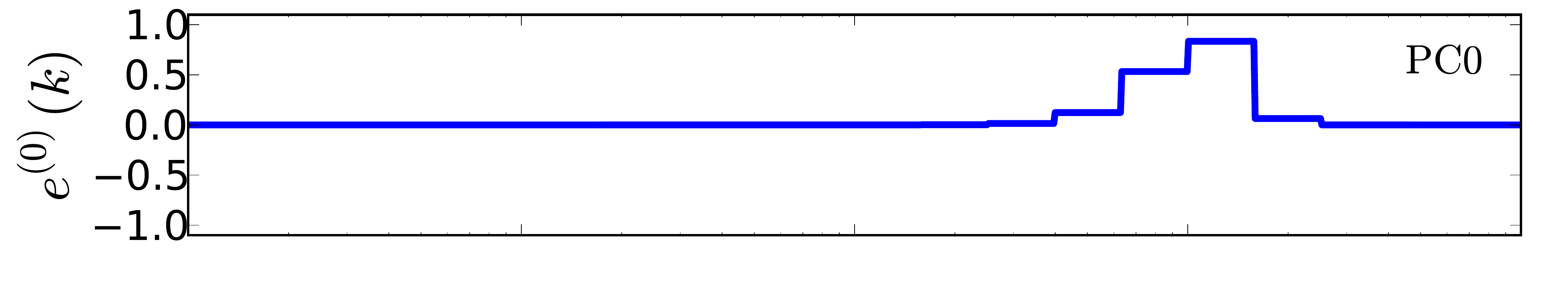} \\[-0.2cm]
\includegraphics[width=2.9in]{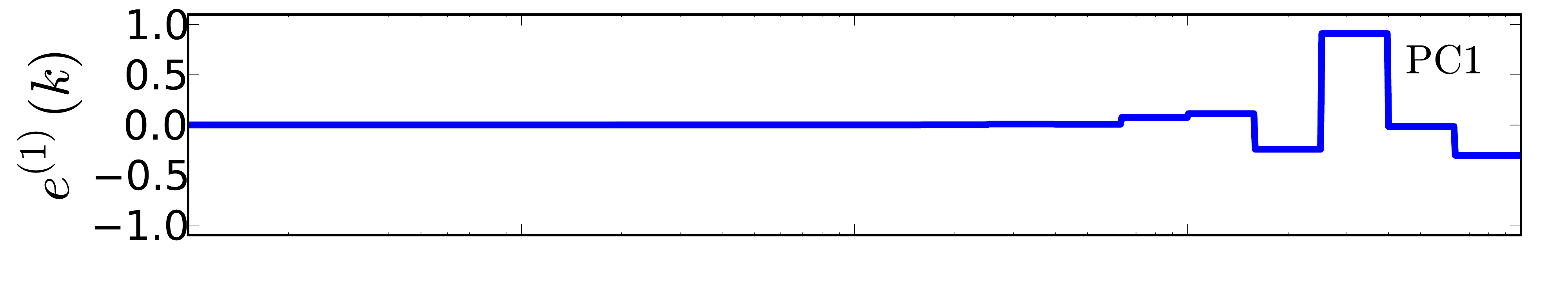} \includegraphics[width=2.9in]{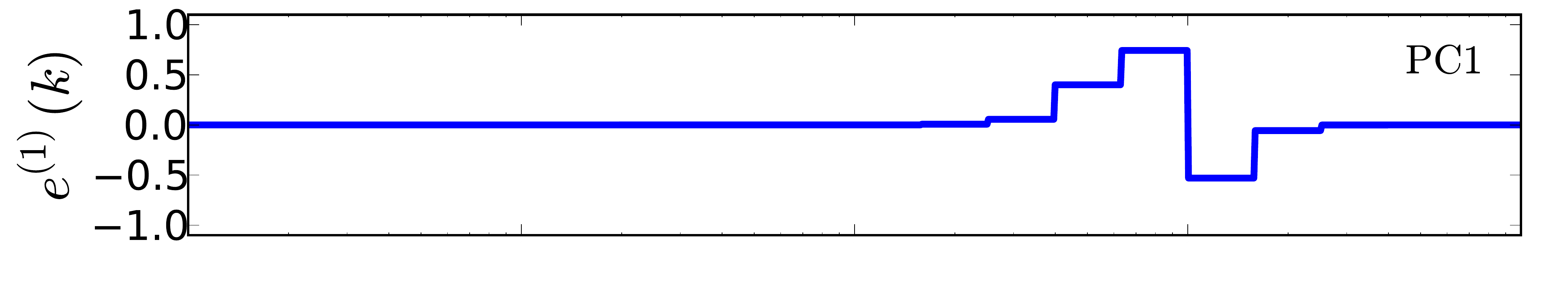} \\[-0.2cm]
\includegraphics[width=2.9in]{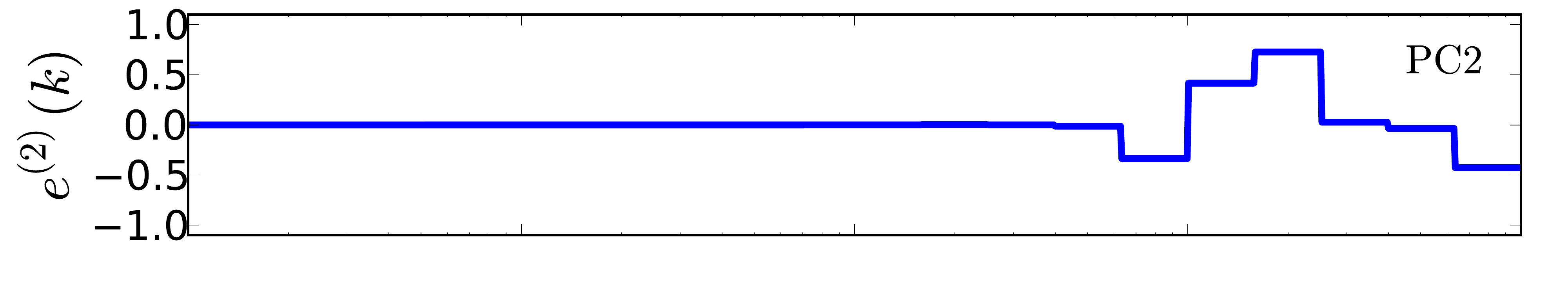} \includegraphics[width=2.9in]{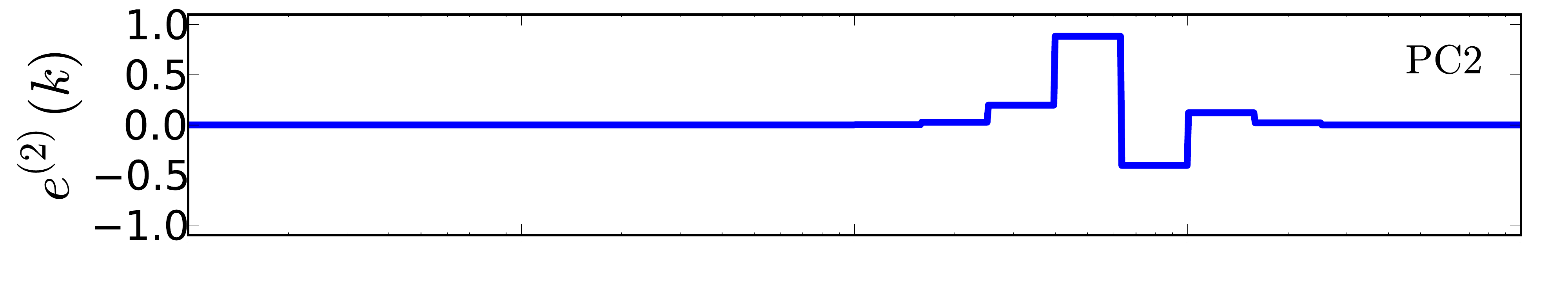} \\[-0.2cm]
\subfigure[BigBOSS PCs]{\includegraphics[width=2.9in]{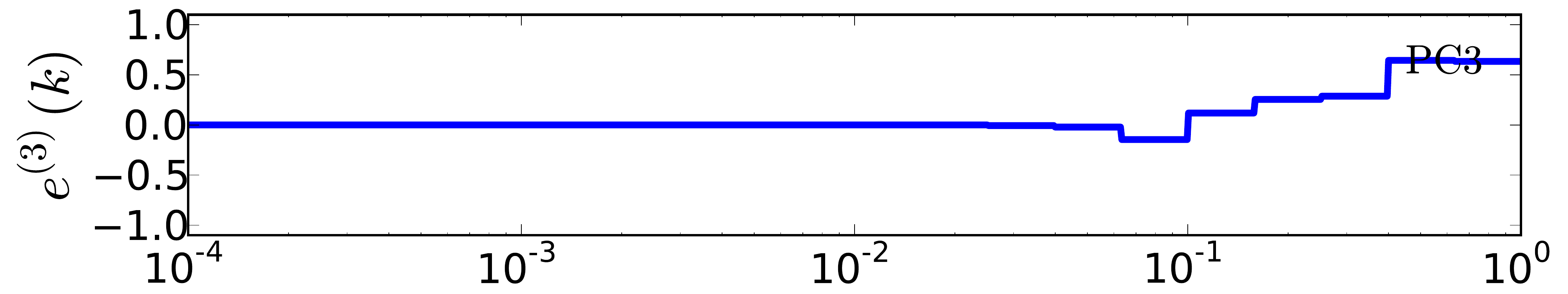}}
\subfigure[Planck PCs]{\includegraphics[width=2.9in]{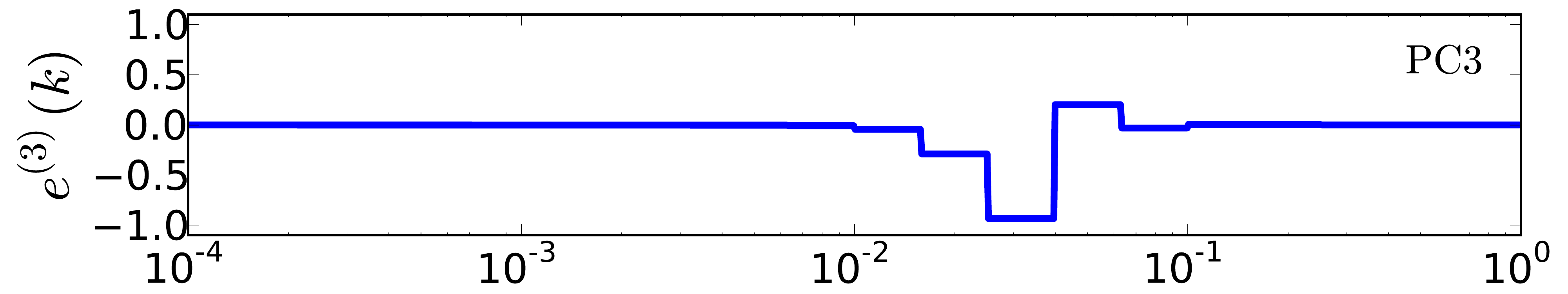}}\\
\includegraphics[width=2.9in]{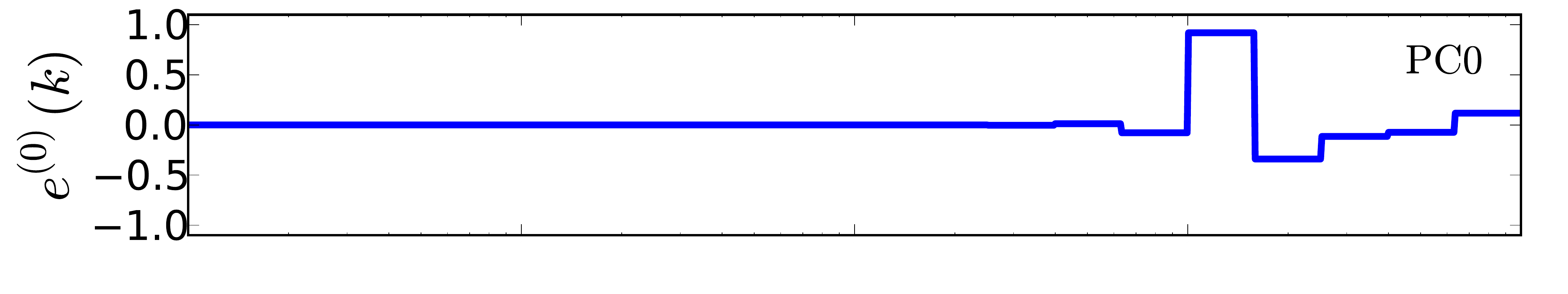} \\[-0.2cm]
\includegraphics[width=2.9in]{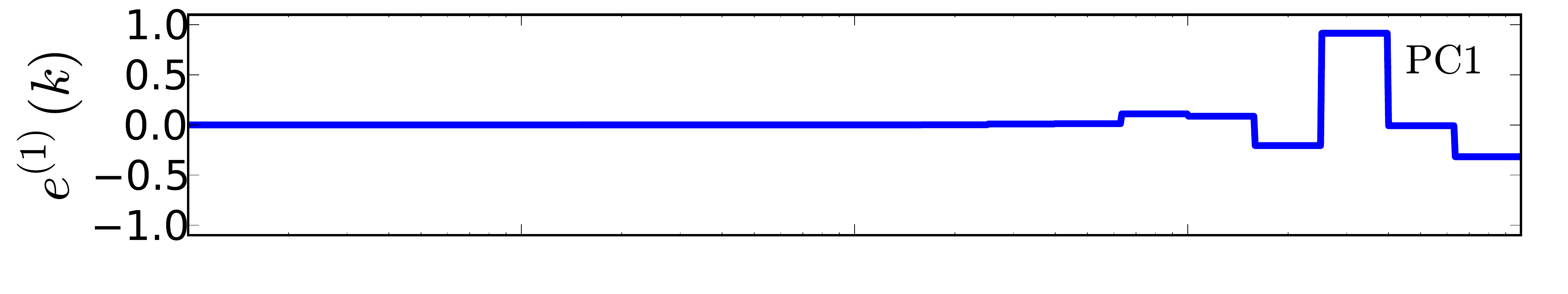} \\[-0.2cm]
\includegraphics[width=2.9in]{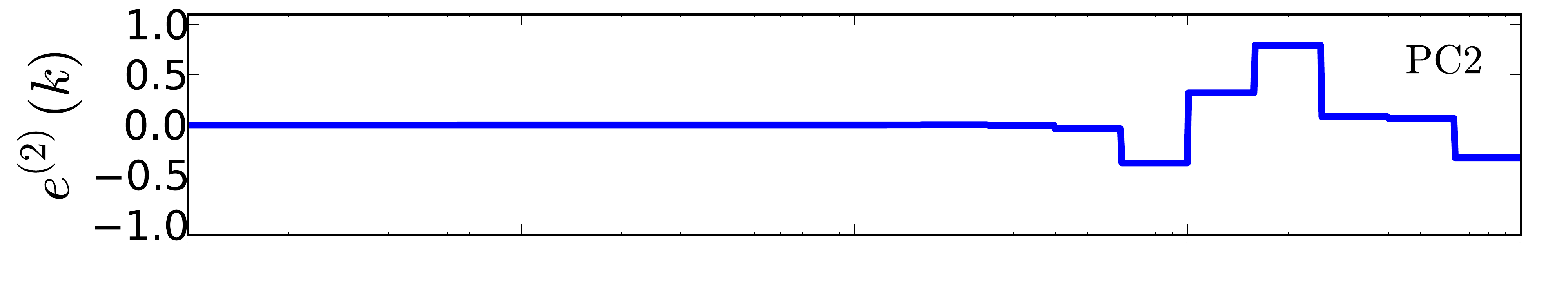} \\[-0.2cm]
\subfigure[Combined PCs]{\includegraphics[width=2.9in]{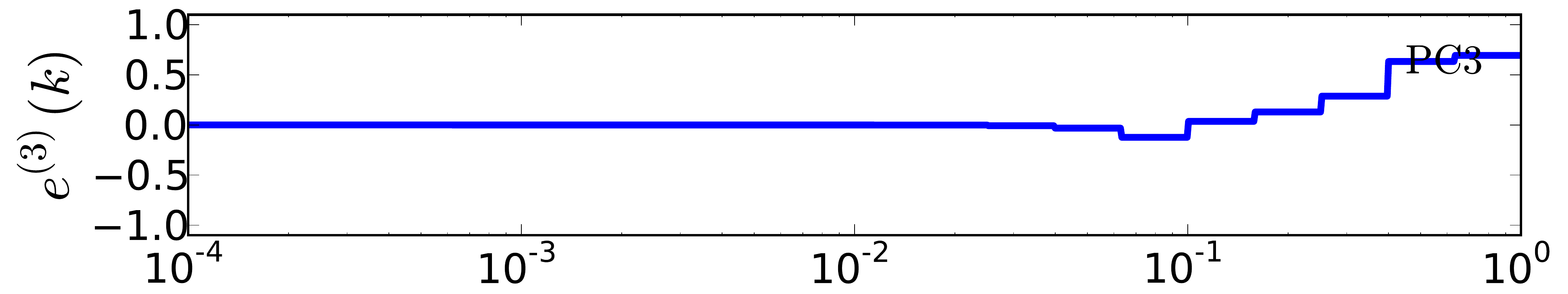}}
\end{center}
\caption{The first four forecasted principal components of $\fnl(k)$ from
  BigBOSS, Planck, and BigBOSS+Planck, assuming the fiducial model $\fnl(k) =
  30$. The PCs eigenvectors $e^{(j)}(k)$ are ordered from the
  best-measured one ($j=0$) to the worst-measured one ($j=19$; not shown here)
  for the assumed fiducial survey. }
\label{fig:lss_cmb_pcs}
\end{figure}

\subsection{Forecasted constraints on the $\fnl^i$}
\label{sec:forecasts}

Figure \ref{fig:unmarg_errs} shows the (unmarginalized) constraints on the
piecewise constant parameters $\fnl^i$ in the generalized local model from
BigBOSS and Planck individually, as well as combined.  Note that the two types
of surveys have comparable constraints at the pivot wavenumber, but the pivot
is at a larger wavenumber for BigBOSS. Away from the pivot, the Planck
constraints are expected to be better than those from BigBOSS, but both
rapidly deteriorate away from their respective pivots. Finally, the combined
constraints are significantly helped by the lever arm in wavenumber when the
two probes are combined, and this leads to better constraints across a wider
range of scales. We will make these statements more quantitative below when we
study the specific case where $\fnl(k)$ is a pure power law in $k$.

The horizontal green curves in all panels in Fig.~\ref{fig:unmarg_errs} show
the accuracy in the constant $\fnl$, projected down from the principal
components $\fnl^i$ as described in BHK11. The accuracy achieved in $\fnl$ is
4.4 for Planck, 2.6 for BigBOSS, and 2.2 for the combined
case.  Recall also that our Fisher matrices for Planck -- but not for BigBOSS
-- assume all cosmological parameters other than the $\fnl^i$ are fixed
(known).

%%%%%%%%%%%%%%%%%%%%%%%%%%%%%%%%%%%%%%%%%%%%%%%
\subsection{Principal Component Analysis}

\begin{figure}[t] %htbp, h = here, t = top, b = bottom, p = page: if nothing there...
\begin{center}
\subfigure[Fiducial $\fnl(k) = 30$, BigBOSS and Planck]{\includegraphics[width= 2.9in]{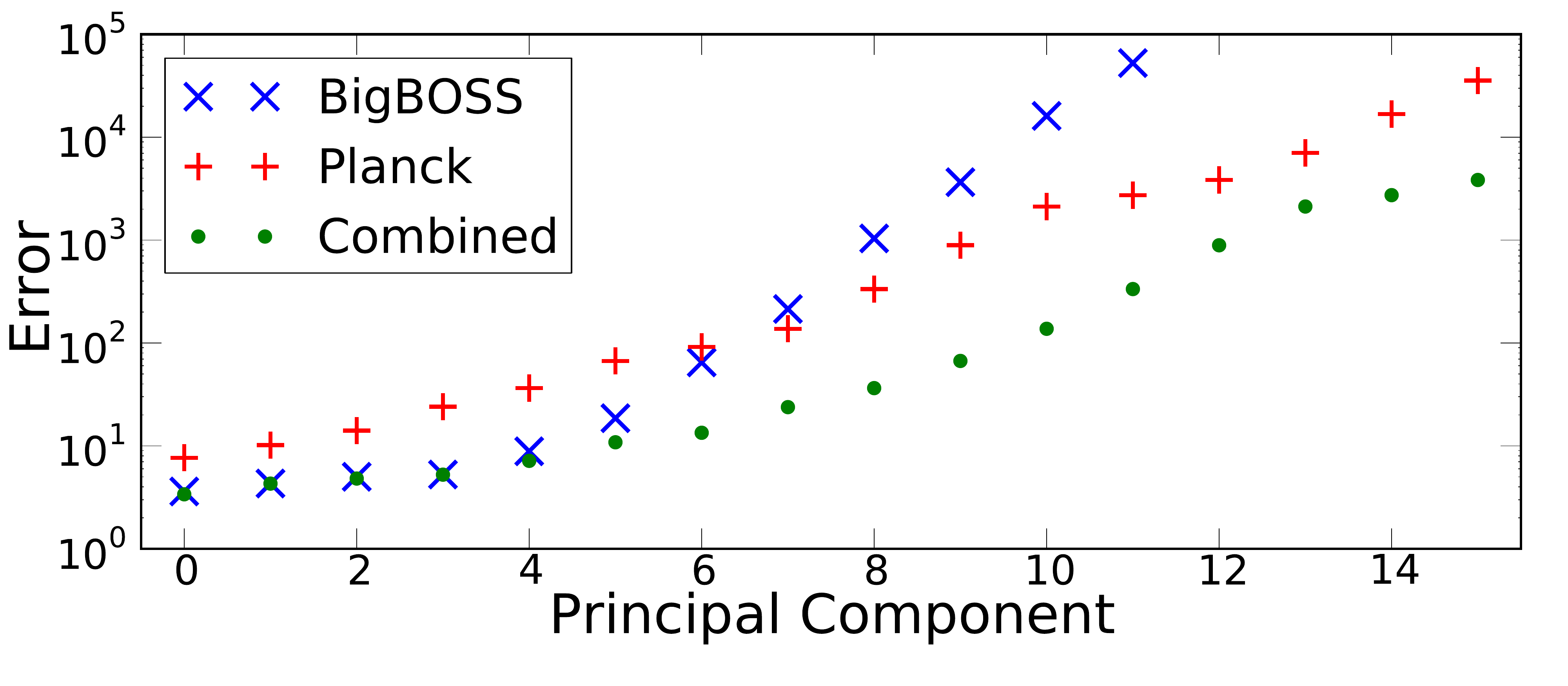}}
\subfigure[Fiducial $\fnl(k) = 0$, BigBOSS and Planck]{\includegraphics[width= 2.9in]{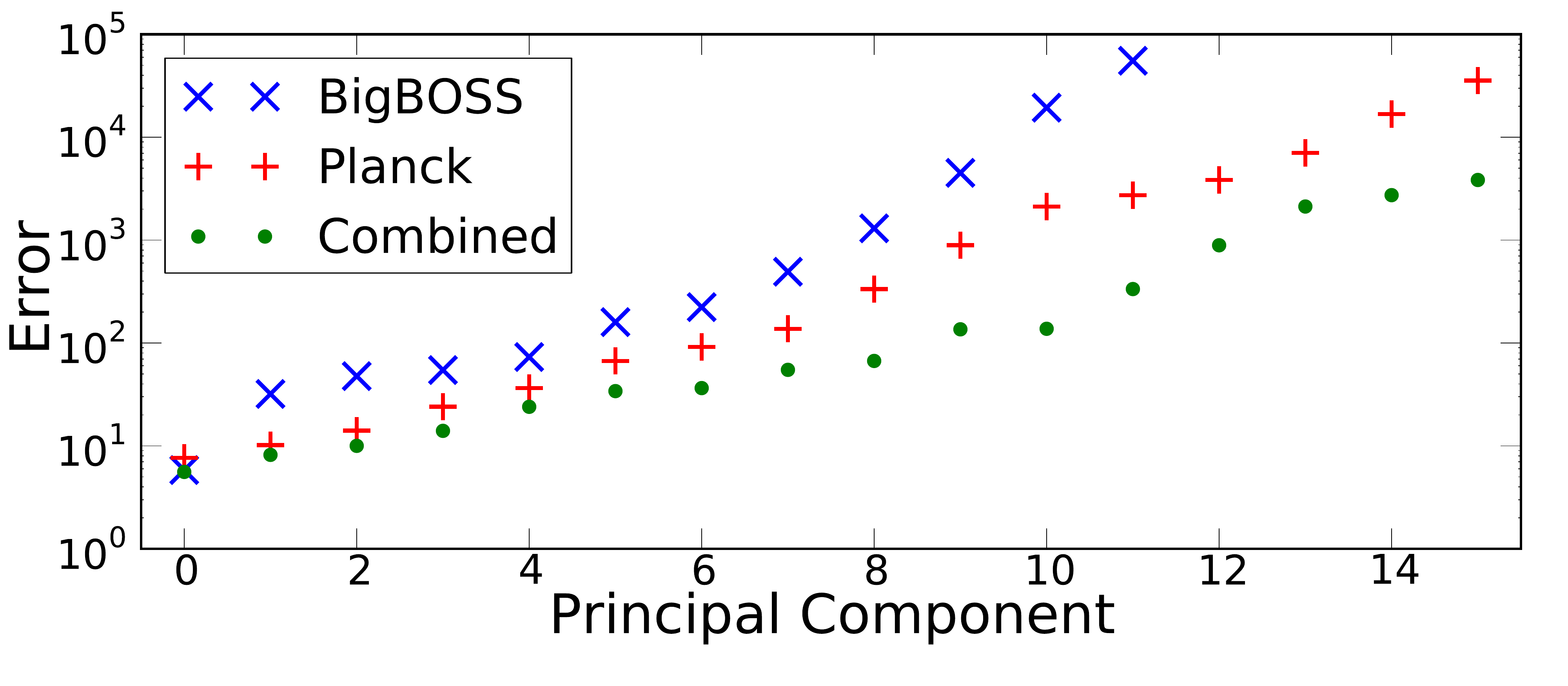}}
%\subfigure[Fiducial $\fnl(k) = 30$, DES and Planck]{\includegraphics[width= 2.9in]{figures/pc_errors_all_des30.pdf}}
        \caption{RMS error on each principal component for BigBOSS, Planck,
          and the two combined. Note that the BigBOSS errors are slightly
          smaller than those from Planck; in all cases combining the CMB and
          LSS decreases errors. Note too that for fiducial $\fnlk=0$ (right
          panel) BigBOSS only constraints $\fnlk$ at the pivot point well (see
          Fig.~\protect\ref{fig:pretty_plot_0}), and hence the error in the
          best-determined principal component is noticeably smaller than
          errors in the other PCs.}
\label{fig:pc_errs}
\end{center}
\end{figure}

Following BHK11, we now represent a general function $\fnl(k)$ in terms of
principal components (PCs). In this approach, the {\it data} determine which
particular modes of $\fnl(k)$ are best or worst measured. The PCs also
constitute a useful form of data compression, so that one can keep only a few
of the best-measured modes to make inferences about the function
$\fnl(k)$. The PCs are weights in wavenumber with amplitudes that are
uncorrelated by construction, and they are ordered from the best-measured
($i=0$) to the worst-measured ($i=19$) for the assumed fiducial survey. We
follow the construction of the PCs following the formalism outlined in
Appendix B of BHK11. We assume a total of 20 principal components distributed
uniformly in $\log(10^{-4}\, \hmpcinv) \leq \log(k)\leq \log(1\, \hmpcinv)$, which is
easily sufficient to make model-independent statements about $\fnl(k)$.

Figure \ref{fig:lss_cmb_pcs} shows the forecasted PCs of LSS and Planck
separately and combined.  Heuristically, the lowest principal component (PC0)
serves to see how well we can find the deviation of $\fnlk$ at its pivot
(i.e.\ best-determined wavenumber) from the fiducial value. The higher PCs
(PC1, PC2, etc) serve to probe the k-dependence of $\fnl$.

Figure \ref{fig:pc_errs} shows the $1$-$\sigma$ errors on the PCs for BigBOSS,
Planck, and the two combined.  Note that the BigBOSS errors are slightly
smaller than those from Planck (the DES errors, not shown here, are bigger
than Planck's). Combining BigBOSS and Planck sharply decreases the
errors. Note too that for fiducial $\fnlk=0$, BigBOSS only constraints $\fnlk$
at the pivot point well (as shown below in Fig.~\ref{fig:pretty_plot_0}), and
hence the error in the best-determined principal component is noticeably
smaller than errors in the other PCs.

The relative strength of the LSS and CMB constraints at their respective
pivot points strongly depend on two factors: volume of the LSS survey and, to
a slightly lesser extent, fiducial (i.e.\ true) value of $\fnlk$ (the CMB
  is not as sensitive to the fiducial value of $\fnlk$). For example, for
$\fnlk=30$ and DES we find that Planck constraints at the CMB pivot are
stronger, while assuming BigBOSS survey we find that the LSS is slightly
stronger at its own pivot. In addition, the CMB typically constrains $\fnlk$
over a wider range of scales than the LSS. 

%%%%%%%%%%%%%%%%%%%%%%%%%%%%%%%%%%%%%%%%%%%%%%%
%\subsection{Projection and Constraints on Other Models of $\fnl(k)$}
\subsection{Projecting constraints on the power-law model of $\fnl(k)$}

Once the Fisher matrix $F$ has been obtained for the set of parameters
$\fnl^i$, it is quite simple to find the best possible constraints on the
$\fnl^i$ that could be obtained from a future galaxy redshift survey. By
projecting this Fisher matrix into another basis%(see Appendix \ref{app:projection})
, it is also possible to find the constraints on any
arbitrary $\fnl(k)$ without calculating a new Fisher matrix from scratch.

Here we will study the popular simple form of non-Gaussianity analogous to the
conventional parameterization of the power spectrum
\cite{Chen2005,LoVerde,Chen:2006nt,Khoury_Piazza,Byrnes_Choi_Hall,Sefusatti2009}
\begin{equation}
\fnl(k) = \fnl^* \left( k \over k_* \right)^{\nfnl},
\label{eq:runningfnl}
\end{equation}
where $k_*$ is an arbitrary fixed parameter, leaving $\fnl^*$ and $\nfnl$
as the parameters of interest in this model. 
%($k_*$ is generally chosen to minimize degeneracy between 
%$\fnl^*$ and $\nfnl$ for the observable of interest.
%We have set $k_* = 0.165 \hmpcinv$, close to the optimal value in our case;
%in CMB analysis, the optimal value is lower, around $0.06 \hmpcinv$. )  
The partial derivatives of our basis of $\fnl^i$ with respect to these
parameters are:
\begin{eqnarray}
{\partial \fnl^i \over \partial \fnl^*} &=& \left( k_i \over k_* \right)^{\nfnl};
\\[0.2cm]
{\partial \fnl^i \over \partial \nfnl}  &=& \fnl^* \left( k_i \over k_* \right)^{\nfnl} \log \left( k_i \over k_* \right),
\label{eq:powerlaw_partials}
\end{eqnarray}
where $k_i$ is the $k$ at the center of the $i$th $k$-bin. Starting in a basis of 20 $\fnl^i$ evenly spaced in log $k$, we project down
to a basis of $\fnl^*$ and $\nfnl$ in order to forecast constraints on the two new
parameters. [Note that $k_*$ is an arbitrarily chosen parameter which differs
in general from the true pivot $\kpiv$ where constraints on $\fnlk$ are the
best. The choice of $k_*$ affects neither the constraints on $\fnlk$
nor the value of $\kpiv$.] 

\begin{figure}[t] %htbp, h = here, t = top, b = bottom, p = page: if nothing there...
\begin{center}
\includegraphics[width= 6in]{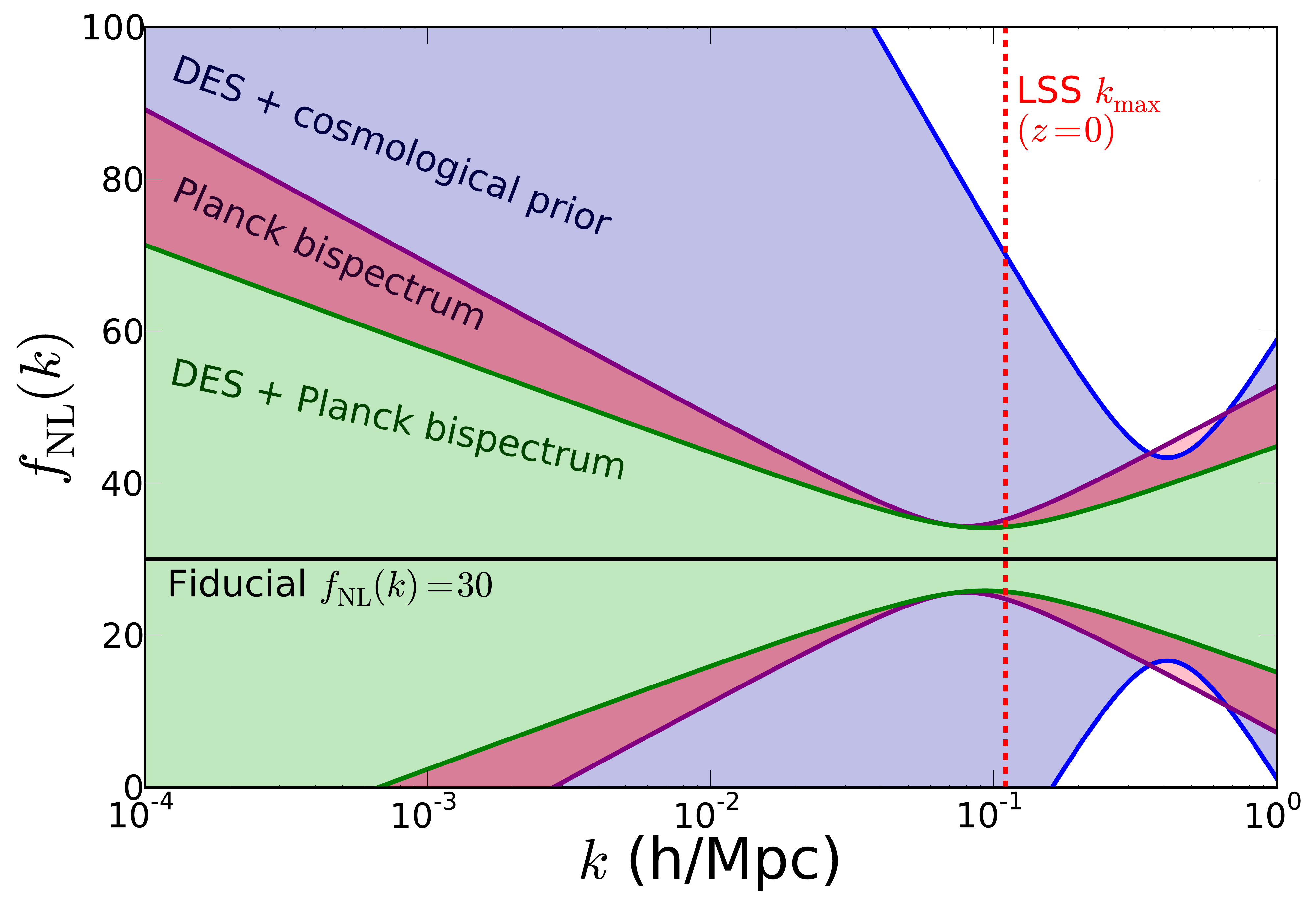}
\caption[Forecasted constraints on the power-law model of $\fnlk$]{Forecasted
  constraints on $\fnl(k)$ from several different data sets, assuming the
  power-law model of scale-dependent non-Gaussianity: $\fnl(k) = \fnl^*
  (k/k_{\rm piv})^{n_{\fnl}}$, projecting down from the piecewise-constant
  $\fnl^i$ basis.  The red dashed line is the maximum $k$ for which
  information was kept in the LSS Fisher matrix at $z = 0$. The LSS survey
  used for this forecast is based on DES.}
\label{fig:pretty_plot_30}
\end{center}
\end{figure}

We can use the constraints on $\fnl^*$ and $\nfnl$ to find constraints on
$\fnlk$ as a whole, through the usual methods of error propagation:
\begin{equation}
\sigma(\fnlk) 
= \sqrt{ \left( {\partial \fnl \over \partial \fnl^*} \sigma(\fnl^*)  \right)^2 
+ \left( {\partial \fnl \over \partial \nfnl} \sigma(\nfnl)  \right)^2
+ 2 {\partial \fnl \over \partial \fnl^*} {\partial \fnl \over \partial \nfnl} C_{\fnl^*, \nfnl}
},
\label{eq:errorprop}
\end{equation}
where $C_{\fnl^*, \nfnl}$ is the covariance between $\fnl^*$ and $\nfnl$, and
$\sigma(\fnl^*)^2$ and $\sigma(\nfnl)^2$ are their respective variances. Using
this relation, and given some fiducial model of $\fnlk$, we can plot the
forecasted constraints on $\fnlk$ as a function of $k$.  This is what we have
done in Figure \ref{fig:pretty_plot_30} for the Planck bispectrum, DES power
spectrum, and the two combined (along with priors on cosmological parameters
from the Planck power spectrum).

\begin{figure}[h] %htbp, h = here, t = top, b = bottom, p = page: if nothing there...
\begin{center}
\includegraphics[width= 6in]{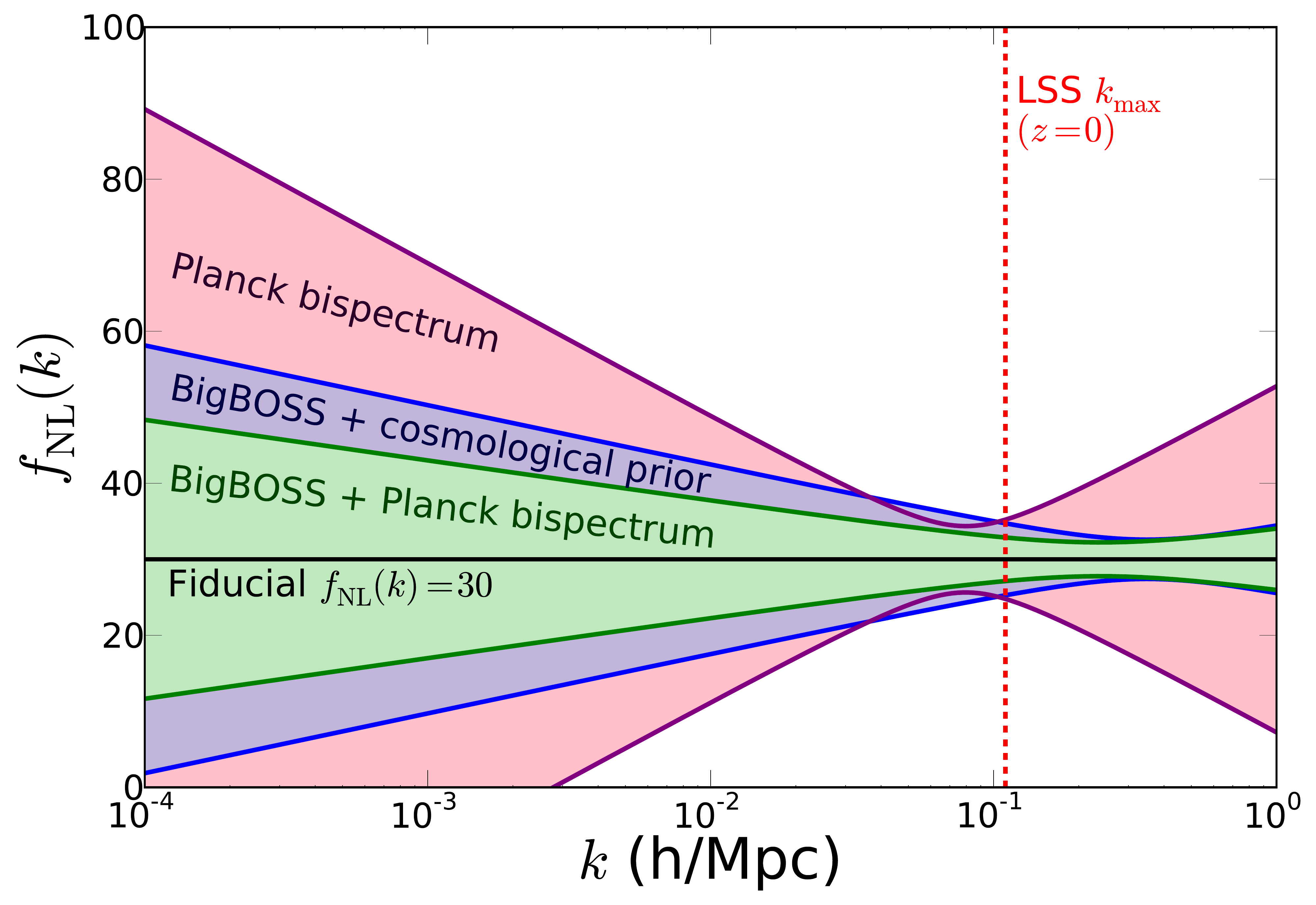}
\caption[The same as Figure \protect\ref{fig:pretty_plot_30}, but with different LSS survey parameters]{The same as Figure \protect\ref{fig:pretty_plot_30}, but with survey parameters for large-scale structure based on BigBOSS.}
% NB: You need to put a \protect on your references in the captions, otherwise LaTeX has a hissy fit and says that there's an "extra }" here.
\label{fig:pretty_plot_BB}
\end{center}
\end{figure}

\begin{table}[b]
\begin{center}
\begin{tabular}{c|c|c|c|c}
\hline\hline         \multicolumn{5}{c}{\rule[-3mm]{0mm}{8mm}Projected errors
  $\sigma(\fnl^*)$ and $\sigma(n_{\fnl})$, and the corresponding pivots}\\
\hline\hline          Variable & \rule[-3mm]{0mm}{8mm} BigBOSS
& \rule[-3mm]{0mm}{8mm} BigBOSS+Planck $C_{\ell}$s
& \rule[-3mm]{0mm}{8mm} Planck bispec
& \rule[-3mm]{0mm}{8mm} {\bf BigBOSS+all Planck} \\\hline\hline
\rule[-2mm]{0mm}{7mm}$\sigma(\fnl^*)$  & 3.0   & 2.6     & 4.4      & 2.2 \\ %\hline
\rule[-3mm]{0mm}{6mm}$\sigma(n_{\fnl})$ & 0.12   & 0.11       & 0.29    & 0.078\\\hline
\rule[-3mm]{0mm}{8mm}  $\fom$ & 2.7  &   3.4     &  0.78   & 5.8  \\\hline
\rule[-3mm]{0mm}{8mm}$k_{piv}$ 	& 0.33 & 0.35   & 0.080  & 0.24 \\\hline \hline
\end{tabular}
\caption{Forecasted constraints on $\fnlstar$ and $\nfnl$ from BigBOSS,
  Planck, and combined data sets for two fiducial values of $\fnlk$.   Each
  column's numbers are for the pivot in that column; thus the errors in the
  two parameters are uncorrelated in each column.  See text for survey specifications.}
\label{tab:running}
\end{center}
\end{table}

Figure \ref{fig:pretty_plot_BB} is analogous to Figure
\ref{fig:pretty_plot_30}, but shows BigBOSS and Planck constraints (rather
than DES and Planck) for the fiducial value of $\fnlk=30$. Note that the
forecasted BigBOSS constraints are, very roughly, comparable to those from
Planck (see also Table \ref{tab:running}), but are also very complementary to
Planck since their best constraints are at a higher $k$.  Our forecasted
constraints on the accuracy of measuring the running with BigBOSS are in good
agreement with forecasts for the Euclid survey in
Ref.~\cite{Giannantonio:2011ya}.

We also introduce the Figure of Merit ($\fom$) of non-Gaussianity. We defined
it analogously to the Figure of Merit for dark energy (\cite{DETF}; see also
\cite{Mort_pcfom}) as 
\begin{equation}
\fom \equiv  (\det F_{2\times 2})^{1/2}
\approx  {6.17 \pi \over A_{95}} 
\label{eq:fom} 
\end{equation}
where $F_{2\times 2}$ is the Fisher matrix projected down to the $2\times 2$ space of
$\fnlstar$ and $\nfnl$, and $A_{95}$ is the area of the 95.4\% confidence
level ellipse in this space. Constraints on the $\fom$ are presented in Table
\ref{tab:running}, and show that combining of BigBOSS and Planck improves
constraints by a factor of between two and five relative to these experiments
alone. What is particularly encouraging is that future constraints will improve
the recently obtained current constraints on the running of non-Gaussianity
\cite{nfnl_wmap7} by more than an order of magnitude.

\begin{table}[t]
\begin{center}
\begin{tabular}{c|c|c|c}
\hline\hline         
\multicolumn{4}{c}{\rule[-3mm]{0mm}{8mm}Projected errors
  $(\sigma_{\fnl^*}, \: \sigma_{n_{\fnl}})$ for different fiducial $\fnlk$}\\
\hline\hline          
%Fiducial $\fnlk$
& \rule[-3mm]{0mm}{8mm} DES
& \rule[-3mm]{0mm}{8mm} BigBOSS
& \rule[-3mm]{0mm}{8mm} Planck
\\\hline\hline
\rule[-3mm]{0mm}{8mm}Fiducial $\fnl(k) = 30$  
& $( 13, \: 1.0)$    
& $( 2.6, \: 0.11)$  
& $( 4.4, \: 0.29)$     
\\ %\hline
\rule[-3mm]{0mm}{7mm}Fiducial $\fnlk = 0$ 
& $(13, \: \infty)$
& $(2.5, \: \infty)$       
& $(4.4, \: \infty)$   
\\\hline \hline
\end{tabular}
\caption[Forecasted constraints on $\fnlstar$ from different LSS surveys,
  assuming different fiducial models, along with forecasted constraints from
  Planck for comparison.]{Forecasted constraints $\sigma_{\fnl^*}$ from
  different LSS surveys, assuming different fiducial models. Forecasted
  constraints from Planck are also shown for comparison. (All values of
  $\nfnl$ are equally likely in the second fiducial model, where $\fnlstar =
  0$.
)}
\label{tab:surveys}
\end{center}
\end{table}

\begin{figure}[htb] %htbp, h = here, t = top, b = bottom, p = page: if nothing there...
\begin{center}
\includegraphics[width= 6in]{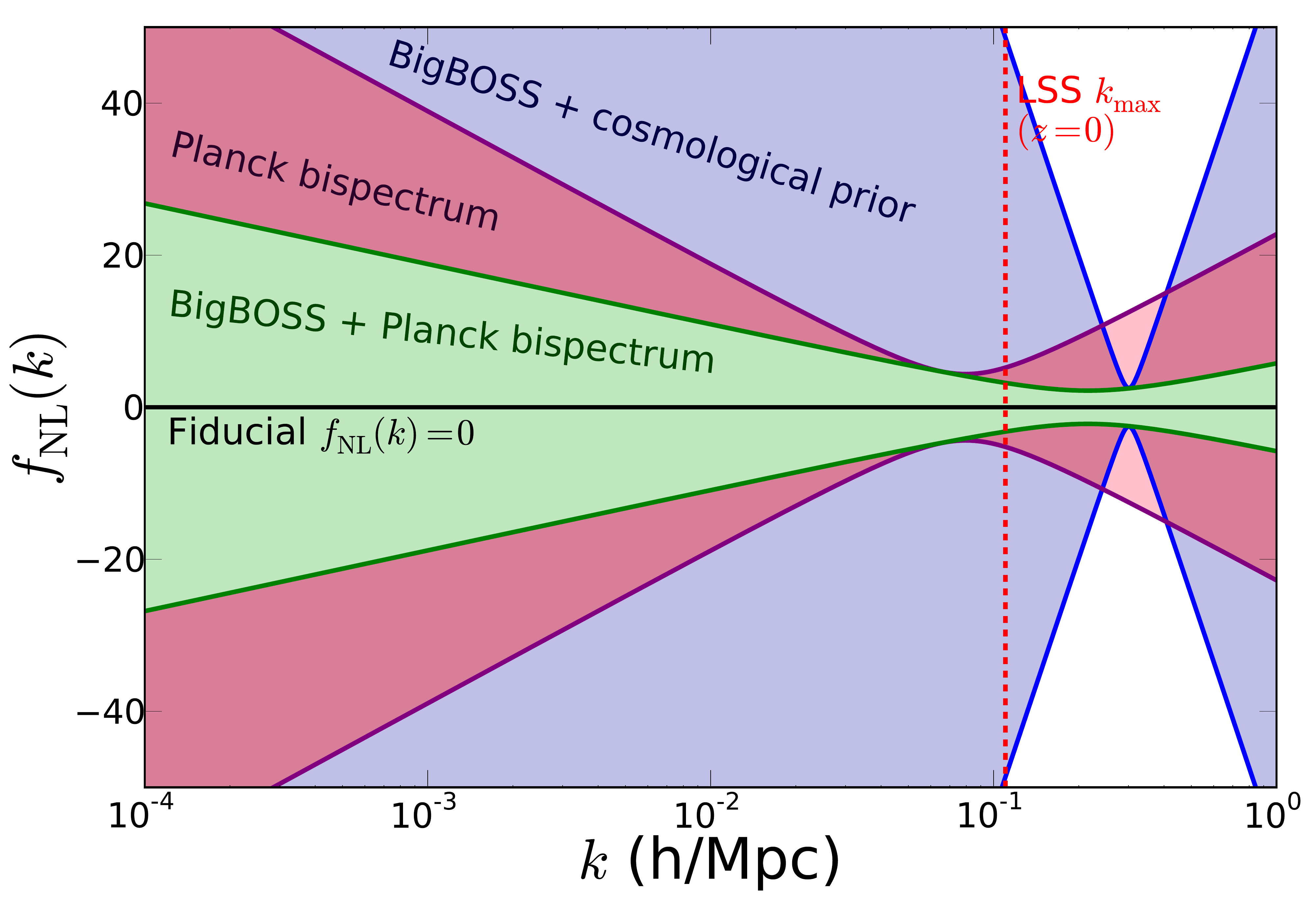}
\caption{ The same as Figure \protect\ref{fig:pretty_plot_BB}, but with a
  fiducial model $\fnlk = 0$. In the limit $\fnl^* \rightarrow 0$ there is no
  information on the running of non-Gaussianity $\nfnl$, and hence the LSS/BigBOSS
  constraints are sharply peaked and essentially constrain $\fnlk$ at only one
  wavenumber.}
% NB: You need to put a \protect on your references in the captions, otherwise LaTeX has a hissy fit and says that there's an "extra }" here.
\label{fig:pretty_plot_0}
\end{center}
\end{figure}

The constraints on $\fnlk$ from a large-scale structure survey are quite
sensitive to the survey parameters. Unlike the constraints on $\fnlk$ from the
CMB bispectrum, the forecasted constraints from LSS are also sensitive to the
choice made for the fiducial model of $\fnlk$, as shown in Section
\ref{sec:fidval}. Forecasted constraints on $\fnlstar$ and $\nfnl$ for the DES
and BigBOSS surveys, with two different fiducial models, are compared to
forecasted constraints from Planck in Table \ref{tab:surveys} (note that {\it
  all} values of $\nfnl$ are equally likely for the fiducial model where
$\fnlstar = 0$, and hence an infinite error on $\nfnl$). 
The scale at which the LSS gives the best constraint (the 'sweet spot')
  turns out to be slightly smaller than the maximum wavenumber assumed to be
  used by the survey, $k_{\rm max}$.
%(which is consistent with the scale with the
%  smallest error from the LSS data as shown in Fig. \ref{fig:unmarg_errs}),
  This is because the halo-bias integration over all the possible momentum
  space configurations in Eq.~(\ref{eq:deltab_b_old}) has dominant
  contributions from small scales\footnote{We performed our bias
      calculations in the Lagrangian picture where the primordial fluctuations
      are linearly extrapolated to $z=0$ as usually done in the
      literature. For an alternative approach including the higher order
      corrections in the framework of the integrated perturbation theory, see
      Ref.~\cite{taka}.}, as we noted previously in \cite{Becker2011}.  Figure
    \ref{fig:pretty_plot_0} shows the same case as
    Fig.~\ref{fig:pretty_plot_BB}, except for the fiducial value of
    $\fnlk=0$. Because $\nfnl$ is arbitrary for this fiducial value,
    constraints on $\fnlk$ are only good at a single, pivot wavenumber; this
    can also be seen by inspection of Eqs.~(\ref{eq:powerlaw_partials}) and
    (\ref{eq:errorprop}). Even in this case (which, note, has measure zero in
    parameter space), we see that combination of BigBOSS and Planck is
    extremely beneficial\footnote{While it may seem surprising that
      constraints away from the pivot wavenumber are finite given that
      $\sigma(\nfnl)=\infty$, we remind the reader that the infinite running
      of $\fnlk$ is essentially multiplied with zero amplitude $\fnlstar$ when
      calculating the constraints at the fiducial value $\fnlk=0$. Closer
      inspection of Eqs.~(\ref{eq:powerlaw_partials}) and (\ref{eq:errorprop})
      confirms this argument.}.

%%%%%%%%%%%%%%%%%%%%%%%%%%%%%%%%%%%%%%%%%%%%%%
\section{Conclusions}\label{sec:concl}

This paper focused on the ability of upcoming LSS and CMB surveys to probe
more general models of primordial non-Gaussianity. We concentrated in
particular on the generalized local model where the parameter $\fnl$ is
promoted to an arbitrary function of scale $\fnl(k)$. Our starting point were the
piecewise constant parameters in $k$, constraints on which are shown in
Fig.~\ref{fig:unmarg_errs}, and their principal components  which are shown in
Fig.~\ref{fig:lss_cmb_pcs} and constrained in Fig.~\ref{fig:pc_errs}. 

Comparison with theory is easiest, however, by using a simpler parametrization
in terms of ``running'' of the spectral index, $\nfnl\equiv  d\ln \fnl(k)/d\ln
k$. Using the two-parameter description of non-Gaussianity in terms of
amplitude $\fnlstar$ and running $\nfnl$, we studied the extent to which a
combination of LSS and CMB observations can constrain the running (Table \ref{tab:running})
and $\fnlk$ as a whole (Figures \ref{fig:pretty_plot_30},
\ref{fig:pretty_plot_BB}, and \ref{fig:pretty_plot_0}).

For the power-law $\fnl(k)$, we found that both the bispectrum measurement
from the CMB Planck survey and power spectrum measurement from an LSS survey
can constrain $\fnlk$ tightly in a relatively narrow range of wavenumbers
around $k\simeq 0.1\hmpcinv$. The scale best constrained by the CMB is larger
(i.e.\ at a smaller $k$) than the scale best constrained by LSS: we get
complementary information about $\fnlk$ from the two data sets.  The ability
of LSS to constrain $\fnlk$ effectively at a wide range of scales depends
on the survey parameters and the fiducial model of $\fnlk$ chosen, as is clear
from Figures \ref{fig:pretty_plot_30}--\ref{fig:pretty_plot_0} and Table
\ref{tab:surveys}. Nonetheless, large galaxy redshift surveys planned for the
future may well be competitive with, or even better than, the constraints on
the magnitude and running of $\fnlk$ expected from Planck.

Beyond the simple power-law model, we find that the combination of CMB and LSS
helps pin down the best-constrained few principal components of $\fnl(k)$
better than either probe alone. Figure \ref{fig:pc_errs} shows that the degree
of complementarity significantly depends on the details of (and systematics
in) the LSS survey.

 The constraints from the DES and BigBOSS, and other upcoming LSS surveys can
 turn out to be worse {\it or better} than those illustrated here, depending
 on how well the systematics can be controlled. While (for example) the
 photometric redshift errors \cite{Cunha_NG}, calibration errors
 \cite{calibration}, and assembly bias of galaxies \cite{Reid_assembly} can all
 introduce parameter biases and degrade constraints, accurate calibration of
 these effects from simulations and observations, as well as selection of the
 ``golden'' class of objects with well understood properties whose clustering
 to use to measure non-Gaussianity, can cancel out these degradations.
 Moreover, we have not considered information from the LSS {\it bispectrum}
 which, while somewhat notoriously difficult to theoretically estimate due to
 non-Gaussian contributions from the gravitational collapse at late times
 (though see \cite{Chan_bisp,Baldauf_tidal} for recent progress on the
 matter), is nevertheless a very potent probe of primordial non-Gaussianity
 (e.g.\ \cite{Sefusatti:2007ih,Jeong_Komatsu_bispec,Sefusatti_halo_bisp,Figueroa}).

Overall, a full exploration of the LSS and CMB systematics is a herculean task 
beyond the scope of this paper; nevertheless, we think  we captured a few key
systematics with our choice of survey specifications and nuisance parameters.
 
Finally, we introduced the figure of merit for measurements of
non-Gaussianity, defined as the inverse area of the constraint region in the
plane of non-Gaussian amplitude and running (see Eq.~(\ref{eq:fom})). We are
very encouraged by the fact that future constraints of non-Gaussianity will
improve current-data figure of merit \cite{nfnl_wmap7} by more than an order of
magnitude, and thus shed interesting constraints on the physics of inflation.

%%%%%%%%%%%%%%%%%%%%%%%%%%%%%%%%%%%%%%%%%%%%%%%
\section{Acknowledgments}
We thank Donghui Jeong and Amit Yadav for helpful feedback during this
project.  We thank the Aspen Center for Physics, which is supported by the
National Science Foundation Grant No.\ 1066293, for the hospitality in the
summer 2010 and 2012 programs. We also acknowledge the use of the publicly
available CAMB \cite{CAMB} package. AB and DH were supported by DOE grant
under contract DE-FG02-95ER40899, NSF under contract AST-0807564, and NASA
under contract NNX09AC89G. KK was supported by MCTP and Ministry of Education,
Sports, Science and Technology (MEXT) of Japan.

%%%%%%%%%%%%%%%%%%%%%%%%%%%%%%%%%%%%%%%%%%%%%%%
\appendix

%%%%%%%%%%%%%%%%%%%%%%%%%%%%%%%%%%%%%%%%%%%%%%%
\section{Calculating the CMB bispectrum}\label{app:bispec}

Calculating the CMB bispectrum is a problem that has been well-studied
elsewhere in the literature, both for the general case and primordial local
non-Gaussianity (e.g.\ \cite{Bartolo:2004if}). Here, we briefly review the
technique for calculating the bispectrum in the case of local non-Gaussianity,
as well as the extension to the generalized local model that we discuss in
this paper.

The bispectrum is defined as:
\begin{equation}
B_{\ell_1 \ell_2 \ell_3, m_1 m_2 m_3} \equiv \langle a_{\ell_1 m_1} a_{\ell_2 m_2} a_{\ell_3 m_3} \rangle
\label{eq:CMBbispec}
\end{equation}
where the $a_{\ell m}$s are the coefficients on the spherical harmonic
decomposition of the CMB sky. The $a_{\ell m}$s can be related to the Bardeen
curvature perturbations $\Phi(\mathbf{k})$ by:
\begin{equation}
a_{\ell m} = \int d^2 \mathbf{\hat{k}} \: {\Delta T (\mathbf{\hat{k}}) \over T} Y^*_{\ell m}(\mathbf{\hat{k}}) =  4 \pi (- i)^\ell \int {d^3 k \over (2 \pi)^3} \: \Phi(\mathbf{k}) g_\ell(k) Y^*_{\ell m}(\mathbf{\hat{k}})
\label{eq:alm}
\end{equation}
Here, $g_\ell (k)$ is the CMB temperature radiation transfer function. There are
several conventions used for this transfer function; $g_\ell (k)$ is related to
the transfer function $T_\ell (k)$ found in (\cite{Gibelyou2010}) by:
\begin{equation}
g_\ell  (k) = {(-i)^\ell  \over \sqrt{2 \ell  (\ell  + 1)}} T_\ell  (k)
\end{equation}
%We will be using yet another convention, as both of the transfer functions
%above lead to messy prefactors later on. 
Throughout this paper, we denote the radiation transfer functions as $t_\ell
(k)$, defined as:
\begin{equation}
t_\ell  (k) = { 1 \over (- i)^\ell } g_\ell  (k) = {1 \over \sqrt{2 \ell  (\ell  + 1)}} T_\ell  (k)
\end{equation}
With these transfer functions, \eqref{eq:alm} becomes:
\begin{equation}
a_{\ell  m} =  {4 \pi \over \sqrt{2 \ell  (\ell  + 1)} } (- 1)^\ell  \int {d^3 k \over (2 \pi)^3} \: \Phi(\mathbf{k}) t_\ell (k) Y^*_{\ell m}(\mathbf{\hat{k}}).
\label{eq:new_alm}
\end{equation}
%One last word on transfer function conventions: these transfer functions
%connect the CMB sky to the Bardeen curvature perturbations, not the primordial
%curvature perturbations.

The angular-averaged bispectrum $B_{\ells}$ is related to the raw bispectrum
$B_{\ells, m_1 m_2 m_3}$ of \eqref{eq:CMBbispec} by the relation:
\begin{equation}
B_{\ells} = \sum_{m_1, m_2, m_3} \binom{\ell_1 \;\; \ell_2 \;\; \ell_3}{m_1 \; m_2 \; m_3} B_{\ells, m_1 m_2 m_3}.
\label{eq:aaBispec}
\end{equation}
Here, $\binom{\ell_1 \;\; \ell_2 \;\; \ell_3}{m_1 \; m_2 \; m_3}$ is the Wigner
$3j$-symbol
%% . This symbol ensures that $\ell_1 + \ell_2 + \ell_3$ is even, $m_1 + m_2 +
%% m_3 = 0$, and the triangle inequality ($| \ell_i - \ell_j | \leq \ell_k \leq \ell_i +
%% \ell_j$) is met for all $i, j, k$.
%
\footnote{ There are some computational difficulties that arise when
  evaluating the $3j$-symbol for high $l_{1, 2, 3}$; see Appendix \ref{app:3j}
  for more on this.  }.
Substituting \eqref{eq:CMBbispec} and \eqref{eq:new_alm} into
\eqref{eq:aaBispec}, we obtain the following expression for the
angular-averaged bispectrum:
\begin{align}
B_{\ell_1 \ell_2 \ell_3} = 
(4 \pi)^3 (-1)^{\ell_1 + \ell_2 + \ell_3} 
\sum_{m_1, m_2, m_3} \binom{\ell_1 \;\; \ell_2 \;\; \ell_3}{m_1 \; m_2 \; m_3}
\int {d^3 k_1 \over (2 \pi)^3}\: {d^3 k_2 \over (2 \pi)^3} \: {d^3 k_3 \over (2 \pi)^3} \nonumber  \\
\times Y^*_{\ell_1 m_1}(\mathbf{\hat{k_1}}) Y^*_{\ell_2 m_2}(\mathbf{\hat{k_2}}) Y^*_{\ell_3 m_3}(\mathbf{\hat{k_3}})
t_{\ell_1}(k_1) t_{\ell_2}(k_2) t_{\ell_3}(k_3) \langle \Phi(\mathbf{k_1})
\Phi(\mathbf{k_2}) \Phi(\mathbf{k_3}) \rangle .
\label{eq:avgCMBbispec}
\end{align}
Using the definition of the Bardeen curvature bispectrum, $B_{\Phi}$,
\begin{equation}
\langle \Phi(\mathbf{k_1}) \Phi(\mathbf{k_2}) \Phi(\mathbf{k_3}) \rangle = 
(2 \pi)^3 \delta(\mathbf{k_1} + \mathbf{k_2} + \mathbf{k_3}) B_{\Phi} (k_1, k_2, k_3),
\label{eq:CurvBispec}
\end{equation}
we find
\begin{align}
B_{\ell_1 \ell_2 \ell_3} = 
{1 \over \pi^3} 
\sum_{m_1, m_2, m_3} \binom{\ell_1 \;\; \ell_2 \;\; \ell_3}{m_1 \; m_2 \; m_3}
\int d^3 k_1\: d^3 k_2 \: d^3 k_3
 Y^*_{\ell_1 m_1}(\mathbf{\hat{k_1}}) Y^*_{\ell_2 m_2}(\mathbf{\hat{k_2}}) Y^*_{\ell_3 m_3} (\mathbf{\hat{k_3}}) \nonumber \\
\times
t_{\ell_1}(k_1) t_{\ell_2}(k_2) t_{\ell_3}(k_3) \delta(\mathbf{k_1} + \mathbf{k_2} + \mathbf{k_3}) B_{\Phi} (k_1, k_2, k_3).
\end{align}
(The prefactor of $(-1)^{\ell_1 + \ell_2 + \ell_3}$ vanished because the
Wigner $3j$-symbol ensures $\ell_1 + \ell_2 + \ell_3$ is even.) Taking
advantage of several identities in \cite{Wang_Kam} (their (12) and (13)), the
orthogonality of the spherical harmonics, and the Gaunt integral identity
(\cite{Komatsu_Spergel}), this becomes:
\begin{align}
B_{\ell_1 \ell_2 \ell_3} = 
\left({2 \over \pi}\right)^3
I_{\ells}
& \int k^2_1 d k_1\: k^2_2 d k_2 \: k^2_3 d k_3  
B_{\Phi} (k_1, k_2, k_3) t_{\ell_1}(k_1) t_{\ell_2}(k_2) t_{\ell_3}(k_3)
\nonumber \\
& \times \int_0^{\infty} r^2 d r \;  j_{\ell_1}(k_1 r)
j_{\ell_2}(k_2 r)
j_{\ell_3}(k_3 r),
\label{eq:GeneralResult}
\end{align}
where $I_{\ells}$ is the Gaunt integral
\begin{equation}
I_{\ells} = \sqrt{(2 \ell_1 + 1) (2 \ell_2 + 1) (2 \ell_3 + 1) \over 4 \pi}
\binom{\ell_1 \; \ell_2 \; \ell_3}{0 \;\; 0 \;\; 0}.
%\label{Gaunt_bispec}
\end{equation}
The real-space integral is now a one-dimensional integral in the spherical
coordinate $r$, starting at our location and ending at infinity. This
real-space coordinate is the difference in the conformal time $\Delta\eta =
\int_{t_e}^{t_0} {dt \over a} = c(\tau_0 - \tau_e)$ between the time when the
CMB was emitted and the present. 
%% Equivalently, it is the difference between
%% the radius of the particle horizon of the observable universe when the CMB was
%% observed and that radius when the CMB was first emitted. Thus, 
Nearly all of the contribution to the integral in $r$ comes from a short
period of time around the surface of last scattering, and there are no
physical contributions beyond $r > r_{\rm max} = \eta_0 = c \tau_0 \approx
14.6$ Gpc. To perform this integral, we sampled it 150 times between
$r_{\rm max}$ and $r_{\rm max} - 2 r_*$, where $r_{\rm max} - r_*$ is the
comoving distance to the surface of last scattering. We also sampled 50 times
between $r_{\rm max} - 2 r_*$ and 0 to capture any impact that late-time
effects might have had. Increasing the sampling rate did not significantly
improve our results.

%Just as we would expect from $r$, this has a value of zero at our current
%location, and increases as we go out towards the surface of last scattering
%and the current particle horizon. We cannot receive any sort of signal from
%beyond the particle horizon, so while this integral technically has an upper
%limit at infinity, it doesn't have any physical contributions from $r >
%r_{\rm max} = \eta_0 = c \tau_0 \approx 14.6$ Gpc.

%%%%%%%%%%%%%%%%%%%%%%%%%%%%%%%%%%%%%%%%%%%%%%%
\subsection{Derivatives with respect to $\fnl$ and $\fnl(k)$}

Using \eqref{eq:GeneralResult} along with \eqref{eq:ConstBispec}, we get the
following expression for the angular-averaged CMB bispectrum in the constant
$\fnl$ case:
\begin{align}
B_{\ell_1 \ell_2 \ell_3} = 
2 \Delta^2_{\phi} f_{\rm NL} \left({2 \over \pi}\right)^3
I_{\ells}
&\int k^2_1 d k_1\: k^2_2 d k_2 \: k^2_3 d k_3 
\left( {1 \over k_1^{3 - (n_s - 1)}  k_2^{3 - (n_s - 1)} } + \mbox{perm.} \right) 
 \nonumber  \\
& \times
t_{\ell_1}(k_1) t_{\ell_2}(k_2) t_{\ell_3}(k_3) \int_0^{\infty} r^2 d r \;  j_{\ell_1}(k_1 r)
j_{\ell_2}(k_2 r)
j_{\ell_3}(k_3 r).
\label{eq:ConstCMBbispec}
\end{align}
Following \cite{Komatsu_Spergel,YadavWandelt2010}
%equations 33 and 34 from \cite{YadavWandelt2010} (where they are
%themselves following \cite{Komatsu_Spergel}, equations 17 and 18), 
we define functions $\alpha_{\ell}(r)$ and $\beta_{\ell}(r)$ to help us
rewrite \eqref{eq:ConstCMBbispec} 
%in a more computationally friendly way.
as
\begin{align}
\alpha_{\ell}(r) & \equiv  {2 \over \pi} \int k^2 t_{\ell}(k) j_{\ell}(kr) dk
\label{eq:alpha}
\\
\beta_{\ell}(r) & \equiv {2 \over \pi} \int k^{-(2-n_s)} t_{\ell}(k) j_{\ell}(kr) dk. 
\label{eq:beta}
\end{align}
Now Eq.~\eqref{eq:ConstCMBbispec} reads
\begin{align}
B_{\ell_1 \ell_2 \ell_3} = 
2 \Delta^2_{\phi} f_{\rm NL}
I_{\ells}
 \int_0^{\infty} r^2 d r \left( \alpha_{\ell_1}(r) \beta_{\ell_2}(r) \beta_{\ell_3}(r) + \mbox{perm.} \right)
\label{eq:SimplerConstCMBbispec}
\end{align}
and hence 
\begin{equation}
{\partial B_{\ell_1 \ell_2 \ell_3} \over \partial \fnl}  = {1 \over \fnl} B_{\ell_1 \ell_2 \ell_3}.
\end{equation}

For the scale-{\it dependent} $\fnl(k)$ case, we use \eqref{eq:fnlkBispec} to
find that the angular-averaged CMB bispectrum is:
\begin{align}
{\partial B_{\ell_1 \ell_2 \ell_3} \over \partial \fnl^i}  & = 
2 \Delta^2_{\phi}
I_{\ells}
\int_0^{\infty} r^2 d r \left( \alpha^i_{\ell_1}(r) \beta_{\ell_2}(r) \beta_{\ell_3}(r) + \mbox{perm.} \right)
\label{eq:SimplerfnlkDeriv}
\end{align}
where
\begin{equation}
\alpha^i_{\ell}(r) \equiv {2 \over \pi} \int_{k_i^{\rm lower}}^{k_i^{\rm upper}} k^2 t_{\ell}(k) j_{\ell}(kr) dk.
\label{eq:alphai}
\end{equation}

%%%%%%%%%%%%%%%%%%%%%%%%%%%%%%%%%%%%%%%%%%%%%%%
\subsection{Polarization and cross-terms}
 The bispectrum for multiple fields is a simple extension of the single field
 case. By analogy with Eqs.~\eqref{eq:CMBbispec} and \eqref{eq:alm}, the
 multiple-field bispectrum is
\begin{equation}
B^{pqr}_{\ell_1 \ell_2 \ell_3, m_1 m_2 m_3} = \langle a^p_{\ell_1 m_1} a^q_{\ell_2 m_2} a^r_{\ell_3 m_3} \rangle,
\label{CMBbispecMF}
\end{equation}
where
\begin{equation}
a^p_{\ell m} =  {4 \pi \over \sqrt{2 \ell (\ell + 1)} } (- 1)^\ell \int {d^3 k \over (2 \pi)^3} \: \Phi(\mathbf{k}) t^p_\ell(k) Y^*_{\ell m}(\mathbf{\hat{k}})
\label{almMF}
\end{equation}
and $t^i_\ell(k)$ is either the temperature or polarization radiation transfer
function. Using these definitions and running through
Eqs.~\eqref{eq:avgCMBbispec} through \eqref{eq:SimplerfnlkDeriv} again, we can
rewrite the bispectrum for multiple fields if we just modify
Eqs.~\eqref{eq:alpha}, \eqref{eq:beta}, and \eqref{eq:alphai} slightly:
\begin{align}
\alpha^p_{\ell}(r) & \equiv  {2 \over \pi} \int k^2 t^p_{\ell}(k) j_{\ell}(kr) dk;
\label{alphaMF}
\\
\beta^p_{\ell}(r) & \equiv {2 \over \pi} \int k^{-(2-n_s)} t^p_{\ell}(k) j_{\ell}(kr) dk;
\label{betaMF}
\\
\alpha^{p,i}_{\ell}(r) & \equiv {2 \over \pi} \int_{k_i^{\rm lower}}^{k_i^{\rm upper}} k^2 t^p_{\ell}(k) j_{\ell}(kr) dk.
\label{alphaiMF}
\end{align}
So for the constant $\fnl$ case, we have
\begin{align}
{\partial B^{pqr}_{\ell_1 \ell_2 \ell_3} \over \partial \fnl}= 
2 \Delta^2_{\phi}
I_{\ells}
\int_0^{\infty} r^2 d r \left( \alpha^p_{\ell_1}(r) 
\beta^q_{\ell_2}(r) \beta^r_{\ell_3}(r) + \mbox{perm.} \right),
\label{ConstCMBbispecMF}
\end{align} 
while for the piecewise-constant $\fnl(k)$ case, we have:
\begin{align}
{\partial B^{pqr}_{\ell_1 \ell_2 \ell_3} \over \partial \fnl^i}  & = 
2 \Delta^2_{\phi}
I_{\ells}
\int_0^{\infty} r^2 d r \left( \alpha^{p, i}_{\ell_1}(r) 
\beta^q_{\ell_2}(r) \beta^r_{\ell_3}(r) + \mbox{perm.} \right).
\label{fnlkDerivMF}
\end{align}
%

%%%%%%%%%%%%%%%%%%%%%%%%%%%%%%%%%%%%%%%%%%%%%%%
\section{The covariance of the bispectrum}
\label{app:covariance}

It is usually a good assumption to consider only the Gaussian contribution to
the covariance of the bispectrum, $\mathbf{C}$. Using Wick's theorem, one
can straightforwardly show (\cite{Liguori_AA, BabichZal2004, SpergelGold1999}):
\begin{equation}
\mathbf{C}_{\ells} = C_{\ell_1}C_{\ell_2}C_{\ell_3}
%\label{eq:covariance_app}
\end{equation}
where
 \begin{equation}
C_{\ell } 
= C^{CV}_{\ell } + \sigma^2_{\ell }  W_\ell  
=  C^{CV}_{\ell } + C^N_\ell ,
\label{Cl2}
\end{equation}
where $C^{CV}_\ell $ is cosmic variance, while $C^N_\ell $ is the variance due
to the noise and beam width in the survey; moreover, $\sigma^2_{\ell }$ is the
variance of the noise in the survey per pixel, and $W_\ell $ is a ``window"
term relating to the survey beam type and width (\cite{CoorayHu1999,
  Knox1995}).
 \footnote{Note that \cite{CoorayHu1999} uses $w^{-1}$ for what we are calling
   $\sigma^2$.}  For an experiment with multiple frequency channels (such as
 Planck or WMAP), the basic form of equation \eqref{Cl2} still holds, but
 finding $C^N_\ell $ is slightly trickier (\cite{CoorayHu1999}):
\begin{equation}
\label{freqnoise}
{1 \over C_\ell ^N} 
= \sum_{\nu} {1 \over C^N_\ell  (\nu)} 
= \sum_{\nu} {1 \over \sigma^2_{\ell }(\nu)  W_\ell (\nu) }.
\end{equation}
For uncorrelated Gaussian noise, $\sigma^2_\ell (\nu) = \sigma^2(\nu)$ is
constant, and we can find its value for a particular experiment -- for
example, the Planck beam width and noise parameters are found in the Planck
mission ``blue book."

We have only been dealing with temperature (TT), but it is not significantly
harder to add in polarization (EE) and cross (TE) terms. The covariance matrix
here is (\cite{Yadav2007, BabichZal2004})

\begin{equation}
(\mathbf{C}^{-1}_{\ells})_{lmn, pqr} = (C^{-1}_{\ell_1})_{lp} (C^{-1}_{\ell_2})_{mq} (C^{-1}_{\ell_3})_{nr},
\label{covMF}
\end{equation}
where
\begin{equation}
C_\ell  =
\begin{pmatrix}
C^{TT}_\ell  && C^{TE}_\ell  \\
C^{TE}_\ell  && C^{EE}_\ell 
\end{pmatrix}
.
\label{CMF}
\end{equation}
Noise is dealt with in the same way as in \eqref{Cl2} for $C^{TT}_\ell $ and
$C_\ell ^{EE}$ in \eqref{CMF}. Assuming that the noise for T and E are
uncorrelated, $\sigma_{TE}^2 = \langle \Delta T \Delta E \rangle = \langle
\Delta T \rangle \langle \Delta E \rangle = 0$, and thus $C_\ell ^{N, TE} = 0$ for
all $\ell $.

%%%%%%%%%%%%%%%%%%%%%%%%%%%%%%%%%%%%%%%%%%%%%%%
\section{The high-peak limit}
\label{app:Desjaqcues}

Desjacques et al.\ \cite{Desjacques2011} have identified
a new term that contributes to the scale-dependent bias due to
non-Gaussianity, which becomes important when the high-peak limit assumption
is relaxed.  This new term successfully explains previously mysterious discrepancies
\cite{Shandera2010} between the theoretical expectation for the
scale-dependent bias and the results of numerical simulations. Physically, the
new term accounts for the scale-dependent mapping between the interval in the
peak height $d\nu$ (which is featured in the peak-background split derivation
of the bias) and mass interval $dM$. 

Moreover, this term is only non-zero for cases when $\fnl\neq {\rm const}$,
and therefore it affects constraints on $\fnlk$ that we study in this paper,
but not the numerous forecasts for constant $\fnl$ studied previously in the
literature.

The new term corresponds to the second term of Eq.~\eqref{eq:Desjacques}
\begin{equation}
N(k) \equiv \frac{d \ln F(k)}{d \ln \sigma_{R}}.
\end{equation}
We can make the evaluation of this term more tractable by using the chain rule
\begin{equation}
N(k) = {\sigma_R \over F(k)} {dF \over dM} \left( d \sigma_R \over d M \right)^{-1}.
\label{eq:newterm_app}
\end{equation}
Now we will need to take the derivative of $N(k)$ with respect to the $\fnl^i$,
for our Fisher matrix.
\begin{align}
{\partial N \over \partial \fnl^i}
&= \sigma_R \left(d \sigma_R \over d M \right)^{-1} {\partial \over \partial \fnl^i} \left [ {1 \over F(k)} {dF \over dM} \right]
\nonumber \\[0.2cm]
&= {\sigma_R \over F} \left(d \sigma_R \over d M \right)^{-1} {\partial \over \partial \fnl^i} 
\left [ {d \over dM} \left( \partial F \over \partial \fnl^i \right) - {1 \over F} {dF \over dM} {\partial F \over \partial \fnl^i} \right].
\label{eq:dnewterm_app}
\end{align}

Equations \eqref{eq:newterm_app} and \eqref{eq:dnewterm_app} are everything we
need to properly account for the new term in our Fisher matrix. Note that
$\sigma_R$ and $d \sigma_R/ dM$ are the only redshift-dependent quantities
in $N(k)$; since their redshift dependence is linear and exactly the same, it cancels
entirely, leaving $N(k)$ independent of $z$.

\begin{figure}[h!] %htbp, h = here, t = top, b = bottom, p = page: if nothing there...
\begin{center}
	\subfigure[Unmarginalized without the new term]{\includegraphics[width= 2.9in]{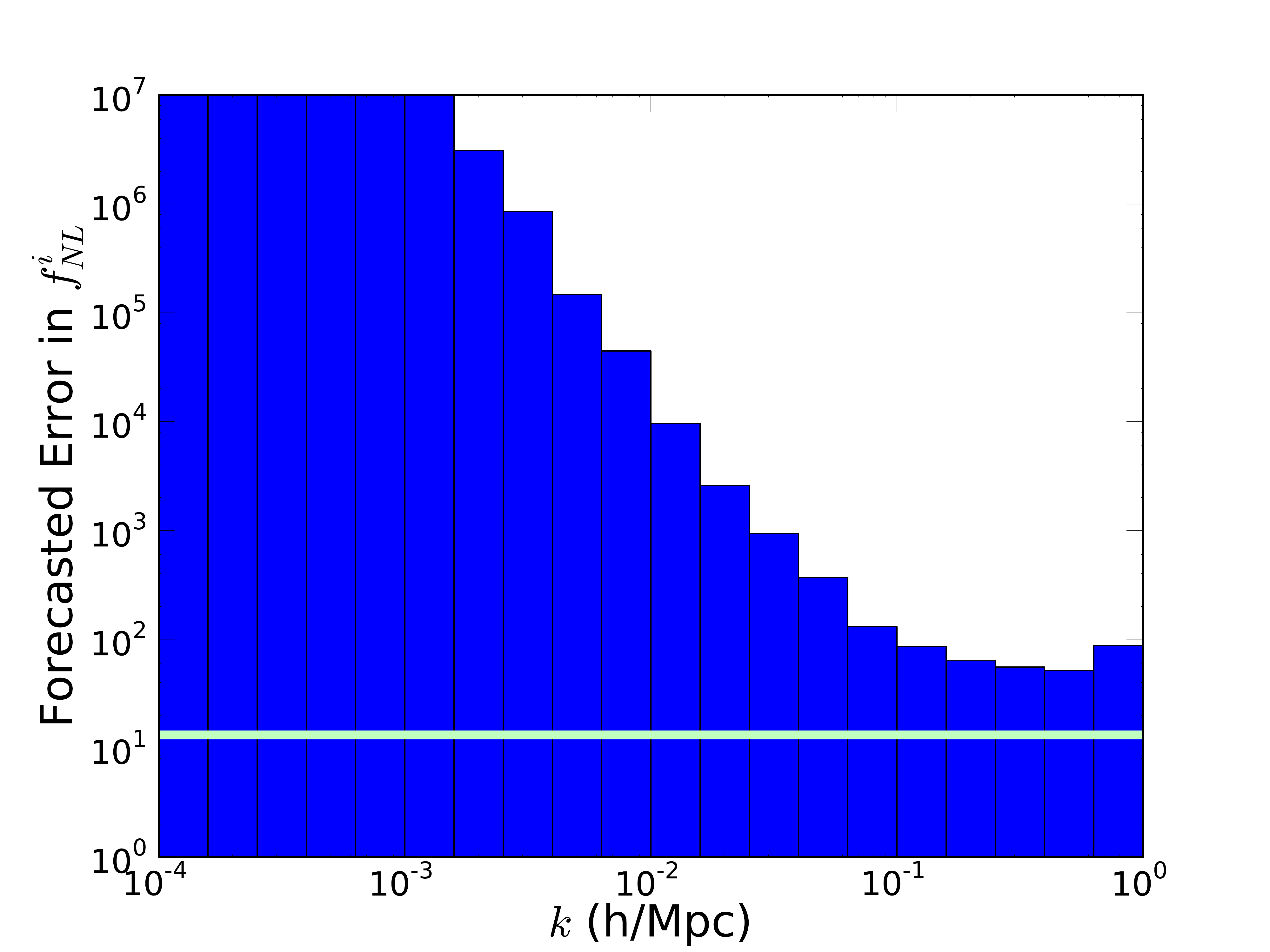}}
	\subfigure[Marginalized without the new term]{\includegraphics[width= 2.9in]{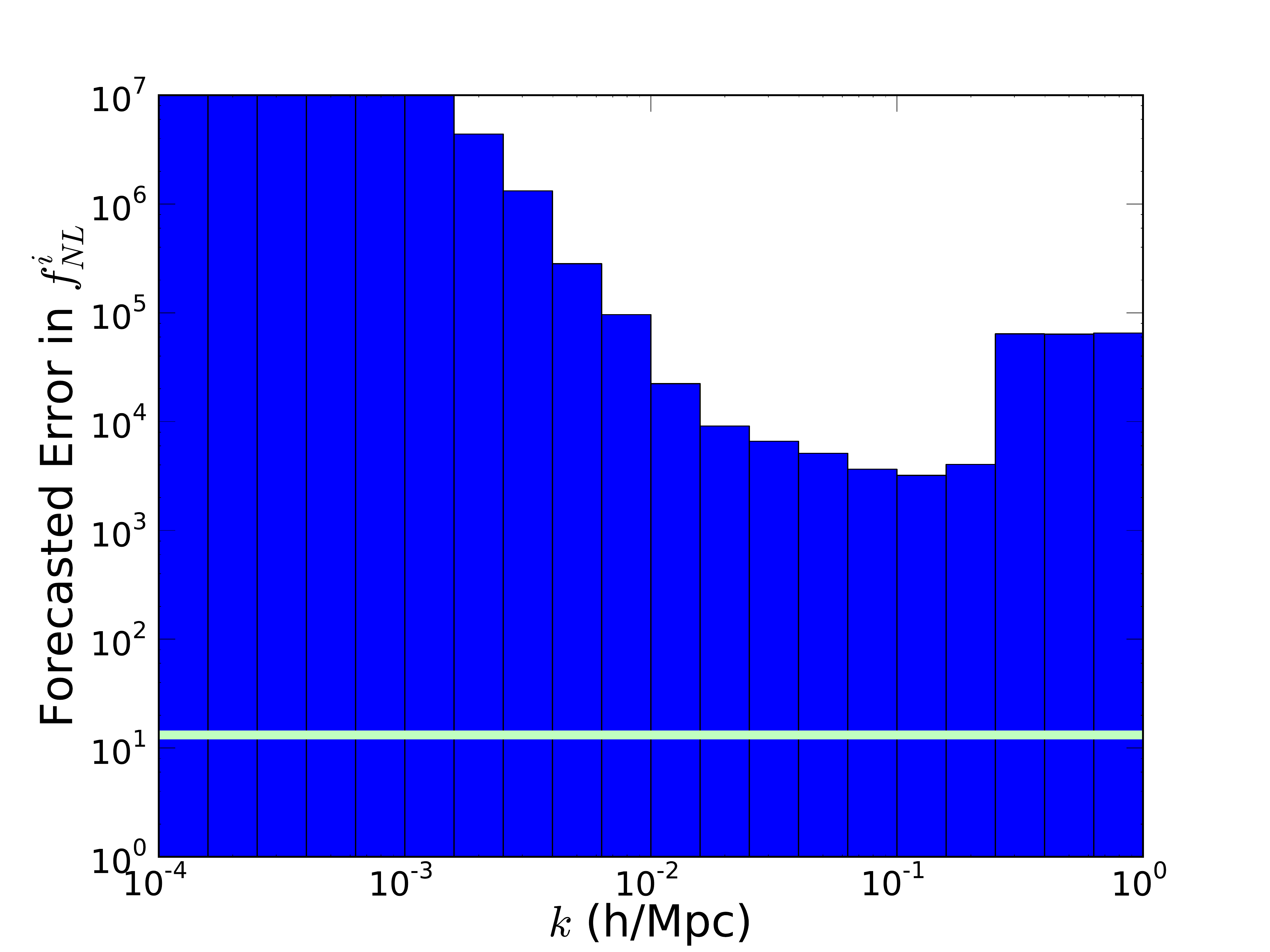}}
	\subfigure[Unmarginalized with the new term]{\includegraphics[width= 2.9in]{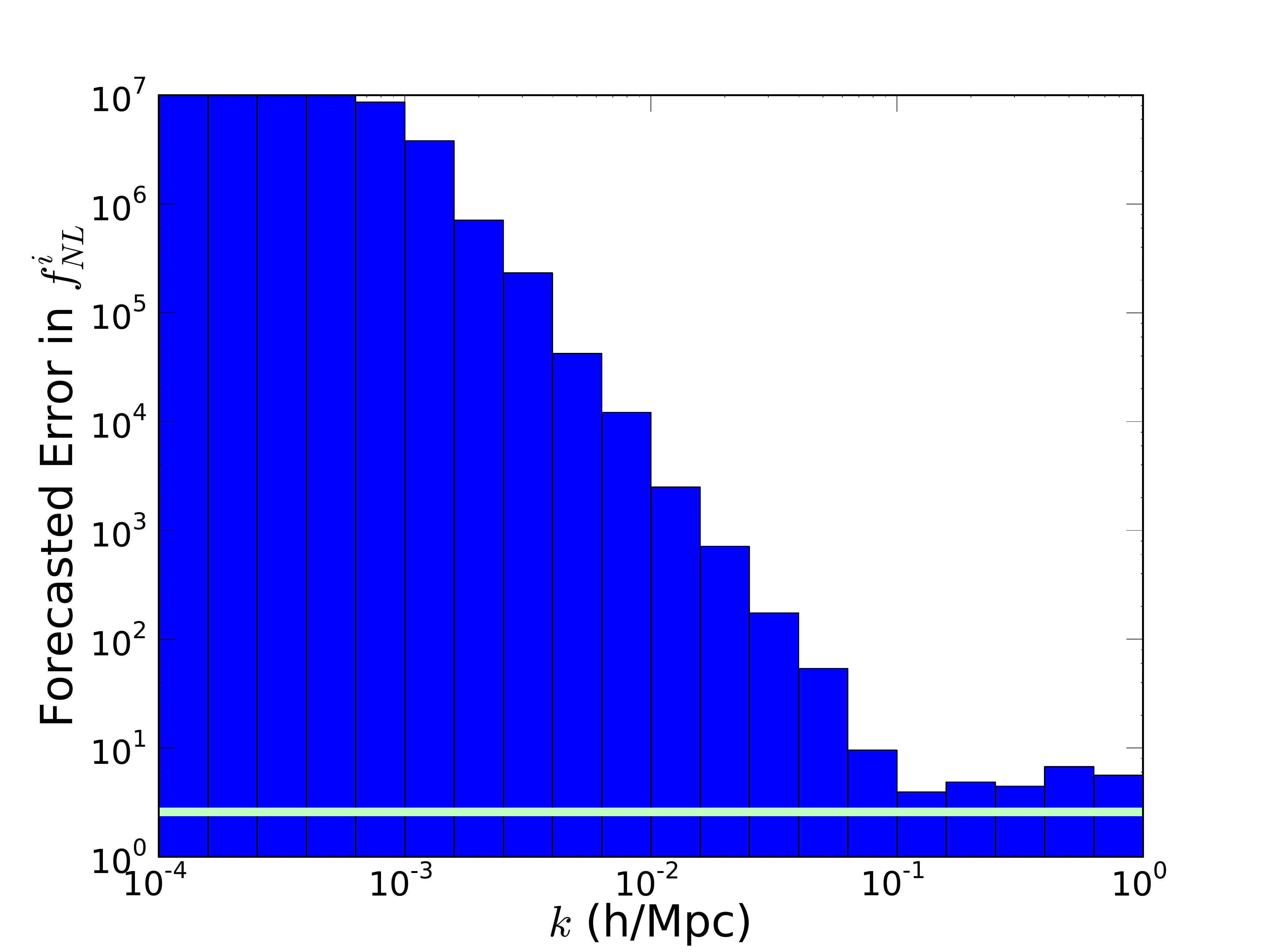}}
	\subfigure[Marginalized with the new term]{\includegraphics[width= 2.9in]{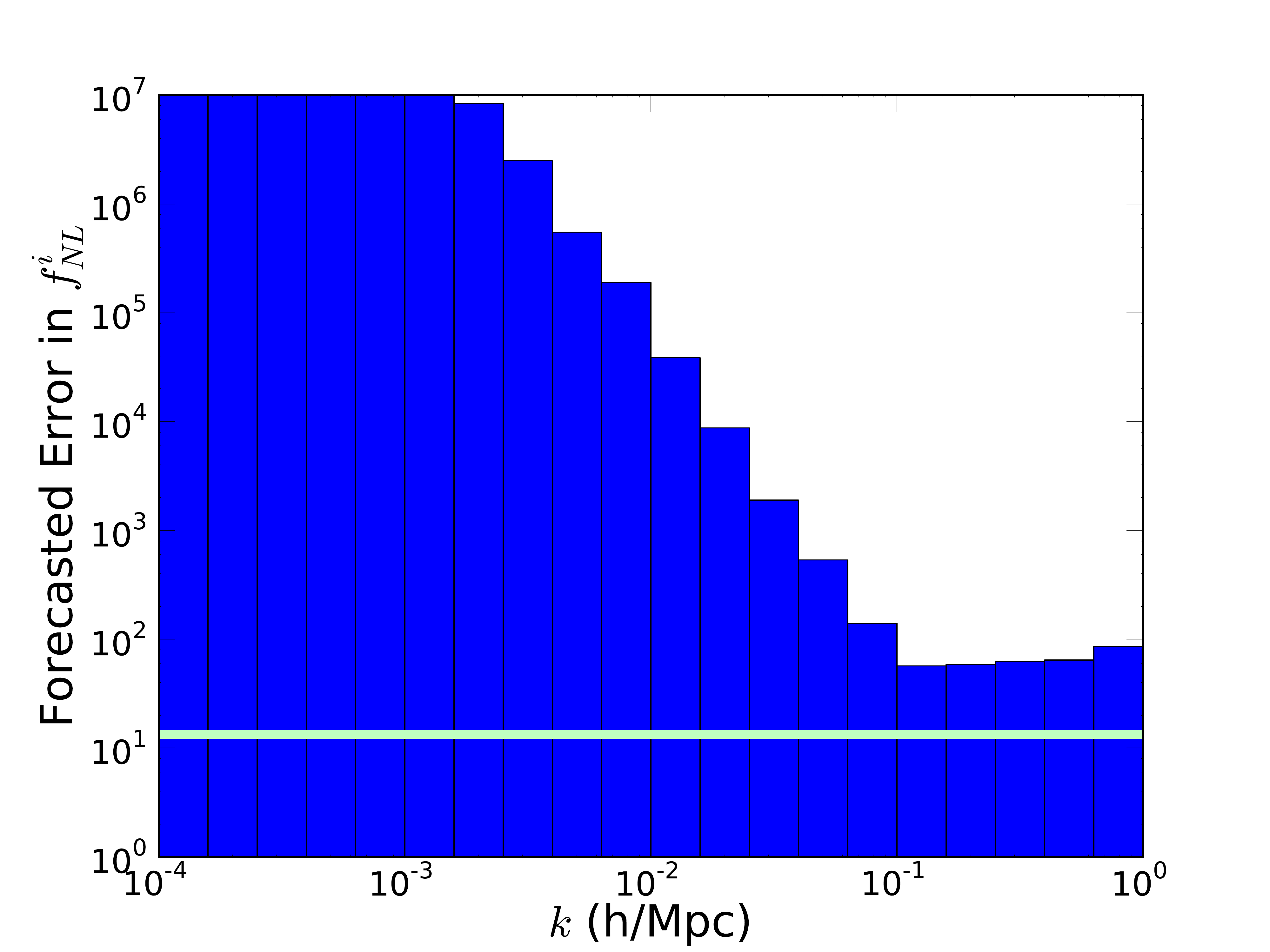}}
        \caption{Illustration of how the inclusion of the correction to
          the scale-dependent bias from \cite{Desjacques2011} affects the
          forecasted constraints on the $\fnl^i$ from DES.  For comparison,
          the green line is the constraint found for a constant $\fnl$ using
          the same assumptions. 
% NB: You need to put a \protect on your references in the captions, otherwise LaTeX has a hissy fit and says that there's an "extra }" here.
}
\label{fig:new_term}
\end{center}
\end{figure}

The effect of this new term on the projected constraints for the $\fnl^i$,
with a fiducial value of $\fnl^i = 30$, are seen in Figure
\ref{fig:new_term}. The figure illustrates that this new term removes much of
the correlation between errors in neighboring $\fnl^i$ and slightly broadens the range of scales
at which the survey is sensitive to $\fnl(k)$. Nevertheless, given that we are
expanding our general $\fnlk$ model around a constant value (30 or zero), the
effects of this new term on the constraints on the amplitude and running of
$\fnl$ -- $\fnl^*$ and $\nfnl$ -- are small.

%%%%%%%%%%%%%%%%%%%%%%%%%%%%%%%%%%%%%%%%%%%%%%%
\section{Calculational  Details}

%%%%%%%%%%%%%%%%%%%%%%%%%%%%%%%%%%%%%%%%%%%%%%%
\subsection{$\ell$ sampling and binning}
In evaluating equation \eqref{eq:fnlkFisher}, we do not actually use every
$\ell \leq \lmax$; that would be incredibly computationally
expensive. Instead, we sample and bin in $\ell$. We keep every $\ell$ up
through $\ell = 40$, at which point sampling drops off gradually until, at
$\ell \gtrsim 100$, only every tenth $\ell$ is sampled. The ``width" of the
bins in $\ell$ are given by the equation
\begin{equation}
\Delta_{\ell_i} = {1 \over 2} \left[ (\ell_i - \ell_{i - 1}) + (\ell_{i + 1} - \ell_i) \right ] = {1 \over 2} (\ell_{i + 1} - \ell_{i - 1}).
\end{equation}

%%%%%%%%%%%%%%%%%%%%%%%%%%%%%%%%%%%%%%%%%%%%%%%
\subsection{Calculating the Wigner $3j$-symbol}
\label{app:3j}
We need to be able to calculate the Wigner $3j$-symbol for large ($ > 1000$)
values of $\ell_{1, 2, 3}$ in order to evaluate many of the expressions we're
interested in. Unfortunately, the $3j$ function built in to the GNU Scientific
Library can't properly evaluate the symbol for $\ell_{1, 2, 3} \gtrsim
70$. Thus, we were forced to create our own special-purpose
$3j$-evaluator. Thankfully, we're only interested in the special case $m_{1,
  2, 3} = 0$; as it turns out, in this case, the $3j$-symbol reduces to (see
Wolfram Mathworld: http://mathworld.wolfram.com/Wigner3j-Symbol.html):
\begin{equation}
\binom{\ell_1 \; \ell_2 \; \ell_3}{0 \;\; 0 \;\; 0}
=
\begin{cases}
\displaystyle
(-1)^g \sqrt{{(2g - 2\ell_1)! (2g - 2\ell_2)! (2g - 2\ell_3)! \over (2g + 1)!}} { g! \over (g - \ell_1)! (g - \ell_2)! (g - \ell_3)!} & \mbox{if } L = 2g; \\[0.3cm]
0 & \mbox{if }  L = 2g + 1,
\end{cases}
\label{3j}
\end{equation}
where $L = \ell_1 + \ell_2 + \ell_3$. Since \eqref{3j} involves evaluating the
factorials of relatively large numbers when any of $l_{1, 2, 3}$ are large, we
used Stirling's approximation to perform the factorials -- but we needed the
factorials to remain accurate even when the arguments were small, so we used
six terms in the approximation.

\bibliographystyle{JHEP}

\bibliography{fnl}

\providecommand{\href}[2]{#2}\begingroup\raggedright\begin{thebibliography}{10}

\bibitem{Chen_AA}
X.~Chen, {\it {Primordial Non-Gaussianities from Inflation Models}},  {\em Adv.
  Astron.} {\bf 2010} (2010) 638979,
  [\href{http://arxiv.org/abs/1002.1416}{{\tt arXiv:1002.1416}}].

\bibitem{Komatsu_CQG}
E.~Komatsu, {\it {Hunting for Primordial Non-Gaussianity in the Cosmic
  Microwave Background}},  {\em Class. Quant. Grav.} {\bf 27} (2010) 124010,
  [\href{http://arxiv.org/abs/1003.6097}{{\tt arXiv:1003.6097}}].

\bibitem{Salopek}
D.~Salopek and J.~Bond, {\it {Nonlinear evolution of long wavelength metric
  fluctuations in inflationary models}},  {\em Phys.Rev.} {\bf D 42} (1990)
  3936--3962.

\bibitem{Verde_CMBLSS}
L.~Verde, L.-M. Wang, A.~Heavens, and M.~Kamionkowski, {\it {Large-scale
  structure, the cosmic microwave background, and primordial non-gaussianity}},
   {\em Mon. Not. Roy. Astron. Soc.} {\bf 313} (2000) L141--L147,
  [\href{http://arxiv.org/abs/astro-ph/9906301}{{\tt astro-ph/9906301}}].

\bibitem{Komatsu_Spergel}
E.~{Komatsu} and D.~N. {Spergel}, {\it {Acoustic signatures in the primary
  microwave background bispectrum}},  {\em Phys. Rev.} {\bf D 63} (Mar., 2001)
  063002, [\href{http://arxiv.org/abs/arXiv:astro-ph/0005036}{{\tt
  arXiv:astro-ph/0005036}}].

\bibitem{Babich_shape}
D.~Babich, P.~Creminelli, and M.~Zaldarriaga, {\it The shape of
  non-gaussianities},  {\em JCAP} {\bf 08} (2004) 009,
  [\href{http://arxiv.org/abs/astro-ph/0405356}{{\tt astro-ph/0405356}}].

\bibitem{Becker2011}
A.~{Becker}, D.~{Huterer}, and K.~{Kadota}, {\it {Scale-dependent
  non-Gaussianity as a generalization of the local model}},  {\em JCAP} {\bf 1}
  (Jan., 2011) 6, [\href{http://arxiv.org/abs/1009.4189}{{\tt
  arXiv:1009.4189}}].

\bibitem{Bean_DBI}
R.~Bean, X.~Chen, H.~Peiris, and J.~Xu, {\it {Comparing Infrared
  Dirac-Born-Infeld Brane Inflation to Observations}},  {\em Phys.Rev.} {\bf
  D77} (2008) 023527, [\href{http://arxiv.org/abs/0710.1812}{{\tt
  arXiv:0710.1812}}].

\bibitem{chris5}
C.~T. Byrnes, M.~Gerstenlauer, S.~Nurmi, G.~Tasinato, and D.~Wands, {\it
  {Scale-dependent non-Gaussianity probes inflationary physics}},  {\em JCAP}
  {\bf 10} (2010) 004, [\href{http://arxiv.org/abs/arXiv:1007.4277}{{\tt
  arXiv:1007.4277}}].

\bibitem{Shandera2010}
S.~Shandera, N.~Dalal, and D.~Huterer, {\it {A generalized local ansatz and its
  effect on halo bias}},  {\em JCAP} {\bf 1103} (2011) 017,
  [\href{http://arxiv.org/abs/1010.3722}{{\tt arXiv:1010.3722}}].

\bibitem{Barnaby_dilaton}
N.~Barnaby, R.~Namba, and M.~Peloso, {\it {Phenomenology of a Pseudo-Scalar
  Inflaton: Naturally Large Nongaussianity}},  {\em JCAP} {\bf 1104} (2011)
  009, [\href{http://arxiv.org/abs/1102.4333}{{\tt arXiv:1102.4333}}].

\bibitem{Planck_paper}
{\it {The Scientific programme of planck}},
  \href{http://arxiv.org/abs/astro-ph/0604069}{{\tt astro-ph/0604069}}.

\bibitem{Sefusatti2009}
E.~{Sefusatti}, M.~{Liguori}, A.~P.~S. {Yadav}, M.~G. {Jackson}, and
  E.~{Pajer}, {\it {Constraining running non-gaussianity}},  {\em JCAP} {\bf
  12} (Dec., 2009) 22, [\href{http://arxiv.org/abs/0906.0232}{{\tt
  arXiv:0906.0232}}].

\bibitem{Giannantonio:2011ya}
T.~{Giannantonio}, C.~{Porciani}, J.~{Carron}, A.~{Amara}, and A.~{Pillepich},
  {\it {Constraining primordial non-Gaussianity with future galaxy surveys}},
  {\em \mnras} {\bf 422} (June, 2012) 2854--2877,
  [\href{http://arxiv.org/abs/1109.0958}{{\tt arXiv:1109.0958}}].

\bibitem{Sefusatti}
E.~Sefusatti, C.~Vale, K.~Kadota, and J.~Frieman, {\it {Primordial
  non-Gaussianity and Dark Energy constraints from Cluster Surveys}},  {\em
  Astrophys.J.} {\bf 658} (2007) 669--679,
  [\href{http://arxiv.org/abs/astro-ph/0609124}{{\tt astro-ph/0609124}}].

\bibitem{Smith_Zaldarriaga}
K.~M. Smith and M.~Zaldarriaga, {\it {Algorithms for bispectra: Forecasting,
  optimal analysis, and simulation}},  {\em Mon.Not.Roy.Astron.Soc.} {\bf 417}
  (2011) 2--19, [\href{http://arxiv.org/abs/astro-ph/0612571}{{\tt
  astro-ph/0612571}}].

\bibitem{Yadav2007}
A.~P. Yadav, E.~Komatsu, and B.~D. Wandelt, {\it Fast estimator of primordial
  non-gaussianity from temperature and polarization anisotropies in the cosmic
  microwave background},  {\em Ap. J.} {\bf 664} (2007) 680.

\bibitem{McDonald}
P.~McDonald, {\it {Primordial non-Gaussianity: large-scale structure signature
  in the perturbative bias model}},  {\em Phys. Rev.} {\bf D78} (2008) 123519,
  [\href{http://arxiv.org/abs/0806.1061}{{\tt arXiv:0806.1061}}].

\bibitem{Carbone}
C.~Carbone, L.~Verde, and S.~Matarrese, {\it {Non-Gaussian halo bias and future
  galaxy surveys}},  {\em Astrophys. J.} {\bf 684} (2008) L1--L4,
  [\href{http://arxiv.org/abs/0806.1950}{{\tt arXiv:0806.1950}}].

\bibitem{Slosar:2008ta}
A.~Slosar, {\it {Optimal dataset combining in $f_{\rm nl}$ constraints from
  large scale structure}},  {\em JCAP} {\bf 0903} (2009) 004,
  [\href{http://arxiv.org/abs/0808.0044}{{\tt arXiv:0808.0044}}].

\bibitem{Carbone2}
C.~Carbone, O.~Mena, and L.~Verde, {\it {Cosmological Parameters Degeneracies
  and Non-Gaussian Halo Bias}},  \href{http://arxiv.org/abs/1003.0456}{{\tt
  arXiv:1003.0456}}.

\bibitem{Fergusson2010}
J.~R. Fergusson, M.~Liguori, and E.~P.~S. Shellard, {\it General cmb and
  primordial bispectrum estimation: Mode expansion, map making, and measures of
  ${F}_{\mathrm{nl}}$},  {\em Phys. Rev. D} {\bf 82} (Jul, 2010) 023502.

\bibitem{Sartoris}
B.~{Sartoris}, S.~{Borgani}, C.~{Fedeli}, S.~{Matarrese}, L.~{Moscardini},
  P.~{Rosati}, and J.~{Weller}, {\it {The potential of X-ray cluster surveys to
  constrain primordial non-Gaussianity}},  {\em \mnras} {\bf 407} (Oct., 2010)
  2339--2354, [\href{http://arxiv.org/abs/1003.0841}{{\tt arXiv:1003.0841}}].

\bibitem{Cunha_NG}
C.~Cunha, D.~Huterer, and O.~Dor\'e, {\it Primordial non-gaussianity from the
  covariance of galaxy cluster counts},  {\em Phys. Rev. D} {\bf 82} (Jul,
  2010) 023004, [\href{http://arxiv.org/abs/1003.2416}{{\tt arXiv:1003.2416}}].

\bibitem{Fedeli:2010ud}
C.~Fedeli, C.~Carbone, L.~Moscardini, and A.~Cimatti, {\it {The clustering of
  galaxies and galaxy clusters: constraints on primordial non-Gaussianity from
  future wide-field surveys}},  {\em Mon.Not.Roy.Astron.Soc.} {\bf 414} (2011)
  1545--1559, [\href{http://arxiv.org/abs/1012.2305}{{\tt arXiv:1012.2305}}].

\bibitem{Joudaki}
S.~Joudaki, O.~Dore, L.~Ferramacho, M.~Kaplinghat, and M.~G. Santos, {\it
  {Primordial non-Gaussianity from the 21 cm Power Spectrum during the Epoch of
  Reionization}},  {\em Phys.Rev.Lett.} {\bf 107} (2011) 131304,
  [\href{http://arxiv.org/abs/1105.1773}{{\tt arXiv:1105.1773}}].

\bibitem{Pillepich:2011zz}
A.~{Pillepich}, C.~{Porciani}, and T.~H. {Reiprich}, {\it {The X-ray cluster
  survey with eRosita: forecasts for cosmology, cluster physics and primordial
  non-Gaussianity}},  {\em \mnras} {\bf 422} (May, 2012) 44--69,
  [\href{http://arxiv.org/abs/1111.6587}{{\tt arXiv:1111.6587}}].

\bibitem{Hazra:2012qz}
D.~K. Hazra and T.~G. Sarkar, {\it {Primordial Non-Gaussianity in the Forest:
  3D Bispectrum of Ly-alpha Flux Spectra Along Multiple Lines of Sight}},
  \href{http://arxiv.org/abs/1205.2790}{{\tt arXiv:1205.2790}}.

\bibitem{Desjacques2011}
V.~Desjacques, D.~Jeong, and F.~Schmidt, {\it Accurate predictions for the
  scale-dependent galaxy bias from primordial non-gaussianity},  {\em Phys.
  Rev. D} {\bf 84} (Sep, 2011) 061301.

\bibitem{Dalal}
N.~Dalal, O.~Dor\'{e}, D.~Huterer, and A.~Shirokov, {\it {The imprints of
  primordial non-gaussianities on large- scale structure: scale dependent bias
  and abundance of virialized objects}},  {\em Phys. Rev.} {\bf D77} (2008)
  123514, [\href{http://arxiv.org/abs/0710.4560}{{\tt arXiv:0710.4560}}].

\bibitem{MV}
S.~Matarrese and L.~Verde, {\it {The effect of primordial non-Gaussianity on
  halo bias}},  {\em Astrophys. J.} {\bf 677} (2008) L77,
  [\href{http://arxiv.org/abs/0801.4826}{{\tt arXiv:0801.4826}}].

\bibitem{Grossi}
M.~{Grossi}, K.~{Dolag}, E.~{Branchini}, S.~{Matarrese}, and L.~{Moscardini},
  {\it {Evolution of Massive Haloes in non-Gaussian Scenarios}},  {\em Mon.
  Not. Roy. Astron. Soc.} {\bf 382} (July, 2007) 1261,
  [\href{http://arxiv.org/abs/0707.2516}{{\tt arXiv:0707.2516}}].

\bibitem{Slosar_etal}
A.~Slosar, C.~Hirata, U.~Seljak, S.~Ho, and N.~Padmanabhan, {\it {Constraints
  on local primordial non-Gaussianity from large scale structure}},  {\em JCAP}
  {\bf 08} (2008) 031, [\href{http://arxiv.org/abs/0805.3580}{{\tt
  arXiv:0805.3580}}].

\bibitem{Afshordi_Tolley}
N.~Afshordi and A.~J. Tolley, {\it {Primordial non-gaussianity, statistics of
  collapsed objects, and the Integrated Sachs-Wolfe effect}},  {\em Phys. Rev.}
  {\bf D78} (2008) 123507, [\href{http://arxiv.org/abs/0806.1046}{{\tt
  arXiv:0806.1046}}].

\bibitem{Taruya08}
A.~{Taruya}, K.~{Koyama}, and T.~{Matsubara}, {\it {Signature of primordial
  non-Gaussianity on the matter power spectrum}},  {\em Phys. Rev. D} {\bf 78}
  (Dec., 2008) 123534, [\href{http://arxiv.org/abs/0808.4085}{{\tt
  arXiv:0808.4085}}].

\bibitem{Desjacques_Seljak_Iliev}
V.~{Desjacques}, U.~{Seljak}, and I.~T. {Iliev}, {\it {Scale-dependent bias
  induced by local non-Gaussianity: a comparison to N-body simulations}},  {\em
  \mnras} {\bf 396} (June, 2009) 85--96,
  [\href{http://arxiv.org/abs/0811.2748}{{\tt arXiv:0811.2748}}].

\bibitem{PPH}
A.~{Pillepich}, C.~{Porciani}, and O.~{Hahn}, {\it {Halo mass function and
  scale-dependent bias from N-body simulations with non-Gaussian initial
  conditions}},  {\em \mnras} {\bf 402} (Feb., 2010) 191--206,
  [\href{http://arxiv.org/abs/0811.4176}{{\tt arXiv:0811.4176}}].

\bibitem{Valageas:2009vn}
P.~Valageas, {\it {Mass function and bias of dark matter halos for non-Gaussian
  initial conditions}},  {\em Astron.Astrophys.} {\bf 514} (2010) A46,
  [\href{http://arxiv.org/abs/0906.1042}{{\tt arXiv:0906.1042}}].

\bibitem{GP}
T.~Giannantonio and C.~Porciani, {\it {Structure formation from non-Gaussian
  initial conditions: multivariate biasing, statistics, and comparison with N-
  body simulations}},  {\em Phys. Rev.} {\bf D81} (2010) 063530,
  [\href{http://arxiv.org/abs/0911.0017}{{\tt arXiv:0911.0017}}].

\bibitem{Schmidt_Kam}
F.~Schmidt and M.~Kamionkowski, {\it {Halo Clustering with Non-Local
  Non-Gaussianity}},  {\em Phys. Rev.} {\bf D 82} (2010) 103002,
  [\href{http://arxiv.org/abs/1008.0638}{{\tt arXiv:1008.0638}}].

\bibitem{Desjacques_long}
V.~Desjacques, D.~Jeong, and F.~Schmidt, {\it {Non-Gaussian Halo Bias
  Re-examined: Mass-dependent Amplitude from the Peak-Background Split and
  Thresholding}},  {\em Phys.Rev.} {\bf D84} (2011) 063512,
  [\href{http://arxiv.org/abs/1105.3628}{{\tt arXiv:1105.3628}}].

\bibitem{Wagner:2010me}
C.~Wagner, L.~Verde, and L.~Boubekeur, {\it {N-body simulations with generic
  non-Gaussian initial conditions I: Power Spectrum and halo mass function}},
  {\em JCAP} {\bf 1010} (2010) 022, [\href{http://arxiv.org/abs/1006.5793}{{\tt
  arXiv:1006.5793}}].

\bibitem{Wagner:2011wx}
C.~Wagner and L.~Verde, {\it {N-body simulations with generic non-Gaussian
  initial conditions II: Halo bias}},  {\em JCAP} {\bf 1203} (2012) 002,
  [\href{http://arxiv.org/abs/1102.3229}{{\tt arXiv:1102.3229}}].

\bibitem{Sefusatti:2011gt}
E.~Sefusatti, M.~Crocce, and V.~Desjacques, {\it {The Halo Bispectrum in N-body
  Simulations with non-Gaussian Initial Conditions}},
  \href{http://arxiv.org/abs/1111.6966}{{\tt arXiv:1111.6966}}.

\bibitem{Grinstein:1986en}
B.~Grinstein and M.~B. Wise, {\it {Nongaussian Fluctuations and the
  Correlations of Galaxies or Rich Clusters of Galaxies}},  {\em Astrophys. J.}
  {\bf 310} (1986) 19--22.

\bibitem{MLB1986}
S.~{Matarrese}, F.~{Lucchin}, and S.~A. {Bonometto}, {\it {A path-integral
  approach to large-scale matter distribution originated by non-Gaussian
  fluctuations}},  {\em Astrophys. J.} {\bf 310} (Nov., 1986) L21--L26.

\bibitem{Tegmark97}
M.~Tegmark, {\it Measuring cosmological parameters with galaxy surveys},  {\em
  Phys. Rev. Lett.} {\bf 79} (1997), no.~20 3806--3809,
  [\href{http://arxiv.org/abs/astro-ph/9706198}{{\tt astro-ph/9706198}}].

\bibitem{SeoEisenstein2003}
H.~{Seo} and D.~J. {Eisenstein}, {\it {Probing Dark Energy with Baryonic
  Acoustic Oscillations from Future Large Galaxy Redshift Surveys}},  {\em
  Astrophys. J.} {\bf 598} (Dec., 2003) 720--740,
  [\href{http://arxiv.org/abs/arXiv:astro-ph/0307460}{{\tt
  arXiv:astro-ph/0307460}}].

\bibitem{BigBOSS_paper}
D.~Schlegel {\em et~al.}, {\it {The BigBOSS Experiment}},
  \href{http://arxiv.org/abs/1106.1706}{{\tt arXiv:1106.1706}}.

\bibitem{wmap7}
E.~Komatsu {\em et~al.}, {\it {Seven-Year Wilkinson Microwave Anisotropy Probe
  (WMAP) Observations: Cosmological Interpretation}},  {\em Astrophys.J.Suppl.}
  {\bf 192} (2011) 18, [\href{http://arxiv.org/abs/1001.4538}{{\tt
  arXiv:1001.4538}}].

\bibitem{BigBOSS_internal}
{\it Bigboss},  tech. rep., 2012.

\bibitem{DES_paper}
T.~Abbott {\em et~al.}, {\it {The dark energy survey}},
  \href{http://arxiv.org/abs/astro-ph/0510346}{{\tt astro-ph/0510346}}.

\bibitem{YadavWandelt2010}
A.~P. Yadav and B.~D. Wandelt, {\it {Primordial Non-Gaussianity in the Cosmic
  Microwave Background}},  {\em Adv.Astron.} {\bf 2010} (2010) 565248,
  [\href{http://arxiv.org/abs/1006.0275}{{\tt arXiv:1006.0275}}].

\bibitem{BabichZal2004}
D.~Babich and M.~Zaldarriaga, {\it {Primordial bispectrum information from CMB
  polarization}},  {\em Phys.Rev.} {\bf D70} (2004) 083005,
  [\href{http://arxiv.org/abs/astro-ph/0408455}{{\tt astro-ph/0408455}}].

\bibitem{SpergelGold1999}
D.~N. Spergel and D.~M. Goldberg, {\it {Microwave background bispectrum. I.
  Basic formalism}},  {\em Phys. Rev.} {\bf D 59} (1999) 103001.

\bibitem{Liguori_Riotto}
M.~Liguori and A.~Riotto, {\it {Impact of Uncertainties in the Cosmological
  Parameters on the Measurement of Primordial non-Gaussianity}},  {\em
  Phys.Rev.} {\bf D78} (2008) 123004,
  [\href{http://arxiv.org/abs/0808.3255}{{\tt arXiv:0808.3255}}].

\bibitem{Chen2005}
X.~Chen, {\it {Running Non-Gaussianities in DBI Inflation}},  {\em Phys. Rev.}
  {\bf D 72} (2005) 123518, [\href{http://arxiv.org/abs/astro-ph/0507053}{{\tt
  astro-ph/0507053}}].

\bibitem{LoVerde}
M.~LoVerde, A.~Miller, S.~Shandera, and L.~Verde, {\it {Effects of
  Scale-Dependent Non-Gaussianity on Cosmological Structures}},  {\em JCAP}
  {\bf 04} (2008) 014, [\href{http://arxiv.org/abs/0711.4126}{{\tt
  arXiv:0711.4126}}].

\bibitem{Chen:2006nt}
X.~Chen, M.-x. Huang, S.~Kachru, and G.~Shiu, {\it Observational signatures and
  non-gaussianities of general single field inflation},  {\em JCAP} {\bf 0701}
  (2007) 002, [\href{http://arxiv.org/abs/hep-th/0605045}{{\tt
  hep-th/0605045}}].

\bibitem{Khoury_Piazza}
J.~Khoury and F.~Piazza, {\it {Rapidly-Varying Speed of Sound, Scale Invariance
  and Non-Gaussian Signatures}},  {\em JCAP} {\bf 0907} (2009) 026,
  [\href{http://arxiv.org/abs/0811.3633}{{\tt arXiv:0811.3633}}].

\bibitem{Byrnes_Choi_Hall}
C.~T. Byrnes, K.-Y. Choi, and L.~M. Hall, {\it {Large non-Gaussianity from
  two-component hybrid inflation}},  {\em JCAP} {\bf 0902} (2009) 017,
  [\href{http://arxiv.org/abs/0812.0807}{{\tt arXiv:0812.0807}}].

\bibitem{DETF}
A.~J. Albrecht {\em et~al.}, {\it {Report of the Dark Energy Task Force}},
  \href{http://arxiv.org/abs/astro-ph/0609591}{{\tt astro-ph/0609591}}.

\bibitem{Mort_pcfom}
M.~J. Mortonson, D.~Huterer, and W.~Hu, {\it {Figures of merit for present and
  future dark energy probes}},  {\em Phys.Rev.} {\bf D82} (2010) 063004,
  [\href{http://arxiv.org/abs/1004.0236}{{\tt arXiv:1004.0236}}].

\bibitem{nfnl_wmap7}
A.~{Becker} and D.~{Huterer}, {\it {First constraints on the running of
  non-Gaussianity}},  \href{http://arxiv.org/abs/(Phys.\ Rev.\ Lett.,
  submitted)}{{\tt (Phys.\ Rev.\ Lett., submitted)}}.

\bibitem{taka}
T.~Matsubara, {\it {Deriving an Accurate Formula of Scale-dependent Bias with
  Primordial Non-Gaussianity: An Application of the Integrated Perturbation
  Theory}},  \href{http://arxiv.org/abs/1206.0562}{{\tt arXiv:1206.0562}}.

\bibitem{calibration}
D.~{Huterer}, C.~{Cunha}, and W.~{Fang}, {\it {Calibration errors unleashed:
  effects on cosmological parameters and requirements for large-scale structure
  surveys}},  \href{http://arxiv.org/abs/(in preparation)}{{\tt (in
  preparation)}}.

\bibitem{Reid_assembly}
B.~A. Reid, L.~Verde, K.~Dolag, S.~Matarrese, and L.~Moscardini, {\it
  {Non-Gaussian halo assembly bias}},  {\em JCAP} {\bf 1007} (2010) 013,
  [\href{http://arxiv.org/abs/1004.1637}{{\tt arXiv:1004.1637}}].

\bibitem{Chan_bisp}
K.~C. Chan, R.~Scoccimarro, and R.~K. Sheth, {\it {Gravity and Large-Scale
  Non-local Bias}},  {\em Phys.Rev.} {\bf D85} (2012) 083509,
  [\href{http://arxiv.org/abs/1201.3614}{{\tt arXiv:1201.3614}}].

\bibitem{Baldauf_tidal}
T.~Baldauf, U.~Seljak, V.~Desjacques, and P.~McDonald, {\it {Evidence for
  Quadratic Tidal Tensor Bias from the Halo Bispectrum}},
  \href{http://arxiv.org/abs/1201.4827}{{\tt arXiv:1201.4827}}.

\bibitem{Sefusatti:2007ih}
E.~Sefusatti and E.~Komatsu, {\it {The bispectrum of galaxies from
  high-redshift galaxy surveys: Primordial non-Gaussianity and non-linear
  galaxy bias}},  {\em Phys.Rev.} {\bf D76} (2007) 083004,
  [\href{http://arxiv.org/abs/0705.0343}{{\tt arXiv:0705.0343}}].

\bibitem{Jeong_Komatsu_bispec}
D.~Jeong and E.~Komatsu, {\it {Primordial non-Gaussianity, scale-dependent
  bias, and the bispectrum of galaxies}},  {\em Astrophys. J.} {\bf 703} (2009)
  1230--1248, [\href{http://arxiv.org/abs/0904.0497}{{\tt arXiv:0904.0497}}].

\bibitem{Sefusatti_halo_bisp}
E.~Sefusatti, M.~Crocce, and V.~Desjacques, {\it {The Halo Bispectrum in N-body
  Simulations with non-Gaussian Initial Conditions}},
  \href{http://arxiv.org/abs/1111.6966}{{\tt arXiv:1111.6966}}.

\bibitem{Figueroa}
D.~Figueroa, E.~Sefusatti, A.~Riotto, and F.~Vernizzi, {\it {The Effect of
  Local non-Gaussianity on the Matter Bispectrum at Small Scales}},
  \href{http://arxiv.org/abs/1205.2015}{{\tt arXiv:1205.2015}}.

\bibitem{CAMB}
A.~Lewis, A.~Challinor, and A.~Lasenby, {\it {Efficient computation of CMB
  anisotropies in closed FRW models}},  {\em Astrophys.J.} {\bf 538} (2000)
  473--476, [\href{http://arxiv.org/abs/astro-ph/9911177}{{\tt
  astro-ph/9911177}}].

\bibitem{Bartolo:2004if}
N.~Bartolo, E.~Komatsu, S.~Matarrese, and A.~Riotto, {\it Non-gaussianity from
  inflation: Theory and observations},  {\em Phys. Rept.} {\bf 402} (2004)
  103--266, [\href{http://arxiv.org/abs/astro-ph/0406398}{{\tt
  astro-ph/0406398}}].

\bibitem{Gibelyou2010}
C.~Gibelyou, D.~Huterer, and W.~Fang, {\it {Detectability of large-scale power
  suppression in the galaxy distribution}},  {\em Phys.Rev.} {\bf D82} (2010)
  123009, [\href{http://arxiv.org/abs/1007.0757}{{\tt arXiv:1007.0757}}].

\bibitem{Wang_Kam}
L.-M. Wang and M.~Kamionkowski, {\it The cosmic microwave background bispectrum
  and inflation},  {\em Phys. Rev.} {\bf D 61} (2000) 063504,
  [\href{http://arxiv.org/abs/astro-ph/9907431}{{\tt astro-ph/9907431}}].

\bibitem{Liguori_AA}
M.~Liguori, E.~Sefusatti, J.~Fergusson, and E.~Shellard, {\it {Primordial
  non-Gaussianity and Bispectrum Measurements in the Cosmic Microwave
  Background and Large-Scale Structure}},  {\em Adv.Astron.} {\bf 2010} (2010)
  980523, [\href{http://arxiv.org/abs/1001.4707}{{\tt arXiv:1001.4707}}].

\bibitem{CoorayHu1999}
A.~R. Cooray and W.~Hu, {\it {Imprint of reionization on the cosmic microwave
  background bispectrum}},  {\em Astrophys.J.} {\bf 534} (2000) 533--550,
  [\href{http://arxiv.org/abs/astro-ph/9910397}{{\tt astro-ph/9910397}}].

\bibitem{Knox1995}
L.~Knox, {\it Determination of inflationary observables by cosmic microwave
  background anisotropy experiments},  {\em Phys. Rev.} {\bf D 52} (1995) 4307.

\end{thebibliography}\endgroup

\end{document}